\documentclass[11pt,preprint]{aastex}
\usepackage{mycommands}
\usepackage{verbatim}
\usepackage{graphicx}
\usepackage[percent]{overpic}
\usepackage{subfigure}
\usepackage{esint}
\usepackage{url}
\begin{document}


\title{Inefficient Angular Momentum Transport in Accretion Disk Boundary Layers: Angular Momentum Belt in the Boundary Layer}

\author{Mikhail A. Belyaev \& Eliot Quataert}
\affil{Astronomy Department, University of California, Berkeley, CA 94720 \\ mbelyaev@berkeley.edu}

\begin{abstract}
We present unstratified 3D MHD simulations of an accretion disk with a boundary layer (BL) that have a duration $\sim 1000$ orbital periods at the inner radius of the accretion disk. We find the surprising result that angular momentum piles up in the boundary layer, which results in a rapidly rotating belt of accreted material at the surface of the star. The angular momentum stored in this belt increases monotonically in time, which implies that angular momentum transport mechanisms in the BL are inefficient and do not couple the accretion disk to the star. This is in spite of the fact that magnetic fields are advected into the BL from the disk and supersonic shear instabilities in the BL excite acoustic waves. In our simulations, these waves only carry a small fraction ($\sim 10 \%$) of the angular momentum required for steady state accretion. Using analytical theory and 2D viscous simulations in the $R-\phi$ plane, we derive an analytical criterion for belt formation to occur in the BL in terms of the ratio of the viscosity in the accretion disk to the viscosity in the BL.   Our MHD simulations have a dimensionless viscosity ($\alpha$) in the BL that is at least a factor of $\sim 100$ smaller than that in the disk.    We discuss the implications of these results for BL dynamics and emission.
\end{abstract}

\section{Introduction}
The transfer of mass from one object to another via accretion is a universal astrophysical process occurring in a wide variety of systems. In this study, we focus on accretion through a gas disk that extends to the surface of a central object with a material outer boundary, such as a white dwarf, neutron star, or protostar, but not a black hole. We assume that the disk extends as a thin disk all the way down to the surface of the compact object (i.e.\ ``star"), which is slowly rotating. This implies that the magnetic field of the accretor is weak enough that the accretion disk is not channeled along magnetic field lines near the surface of the star \citep{GhoshLamb}. 

In our setup, a boundary layer (BL) is present at the interface between the accretion disk and the star. This is the region where the angular velocity of the accretion flow transitions from its nearly Keplerian value in the disk to a much lower value in the star. The BL is energetically important, because for a star rotating below breakup in steady state, about as much energy per unit time must be dissipated in the BL as in the accretion disk proper. This is because half of the gravitational potential energy tapped by accretion goes into rotational kinetic energy at the surface of the star for a Keplerian disk. 

Semi-analytical one-dimensional models were developed to describe the radial structure of and thermal emission from BLs \citep{PringleBL,PringleSavonije,PophamNarayan}. Two-dimensional investigations of the BL in the $r-\theta$ plane have also been undertaken using computer simulations \citep{Kley,Balsara,HertfelderKley1}. All of these models assume that a turbulent viscosity \citep{ShakuraSunyaev,LyndenBellPringle} efficiently couples the disk and the star, decelerating accreted material in the BL. 

In the accretion disk proper, the magnetorotational instability (MRI) is thought to generate the turbulent viscosity that allows material to accrete inward \citep{BalbusMRI}. However, the BL has a rising rotation profile ($d\Omega/dR > 0$) and hence is linearly stable to the MRI. Hence, a different type of instability is likely required to transport angular momentum there. 

Because of the narrow radial extent of the BL, it is natural to consider shear as the physical source of instability within it. However, because the azimuthal flow of material over the surface of the star is highly supersonic\footnote{The jump in azimuthal velocity across the BL is much greater than the sound speed.}, the Kelvin-Helmholtz instability (KHI) does not operate there (at least globally over the entire BL). Rather, shear-acoustic instabilities are excited in the BL \citep{BR}. This class of instabilities was studied in the astrophysical context by \cite{Drury,PPI}, who were interested in their applications to accretion disks. However, shear instabilities are excited on a much shorter timescale in the BL than in the accretion disk proper, because their growth rate is proportional to the shear, and the shear is much greater in the BL than in the accretion disk.

Since shear-acoustic instabilities excite sound waves, \cite{BRS1} proposed that angular momentum transport in the BL is mediated by waves rather than by a turbulent viscosity. Waves are a nonlocal form of angular momentum transport, since they can travel large distances between where they are excited and where they are absorbed. This is in direct contrast to turbulent viscosity, which is a purely local mechanism of angular momentum transport. \cite{BRS2} showed using 3D magnetohydrodynamical (MHD) simulations that shear-acoustic instabilities in the BL can coexist with MRI turbulence in the disk. \cite{HertfelderKley} found them to be present in 2D hydro simulations ($R-\phi$ plane) with radiative transport, and \cite{PhilippovRafikov} studied excitation of shear-acoustic waves in a spreading layer geometry \citep{InogamovSunyaev}.

In this paper, we perform 3D MHD simulations that include a star, disk, and a boundary layer. The focus of our work is on understanding the long term evolution of the system, and we run for $\sim 1000$ orbital periods at the inner edge of the accretion disk. We find that on these long timescales angular momentum piles up in a belt inside the BL region. Moreover, this pile up proceeds for the entirety of the simulation and no steady state is reached. This is in spite of the fact that shear-acoustic instabilities are excited in the BL and persist for the duration of the simulation. This suggests that shear-acoustic instabilities are less efficient at angular momentum transport than previously thought.

In order to understand the reason for the formation of the belt, we also carry out 2D viscous simulations in the $R-\phi$ plane. In our model setup, the viscosity takes different (lower) values in the star and the boundary layer compared to the accretion disk. Because we know the steady state value of the accretion rate through the disk in the viscous simulations, we are able to determine that acoustic waves carry only a small fraction ($\lesssim 10\%$) of the steady state angular momentum current through the disk into the star. This inefficient transport leads to the formation of the rapidly rotating belt on the surface of the star. Additionally, we derive a condition on how small the viscosity in the BL should be relative to the viscosity in the accretion disk for an angular momentum belt to be present there. We also derive the peak amplitude of the angular momentum in the belt in steady state for a given value of the viscosity in the BL. This allow us to connect the viscous simulations with the 3D MHD simulations and derive an upper bound for the BL viscosity in the latter. This upper bound is interesting, because it is much smaller than what is typically assumed in phenomenological models of the BL.

The paper is organized as follows. In \S \ref{sec:3DMHD}, we present the equations and physical setup used in our 3D MHD unstratified simulations. We also check the validity of these simulations by showing that the MRI is resolved in the disk and that shear-acoustic modes are excited in the BL. In \S \ref{sec:belt}, we present results showing the belt of accreted angular momentum in the BL which grows monotonically in time without bound in the 3D MHD simulations. In \S \ref{sec:visc}, we present the results of 2D viscous hydro simulations. Additionally, we use viscous theory to derive a physical criterion that must be met for the angular momentum belt to form in the BL. We also provide an estimate of the peak angular momentum in the belt and derive an upper bound for the BL viscosity in the 3D MHD simulations. \S \ref{sec:discuss} discusses the implications of our work for BL dynamics and emission. 

\section{3D Unstratified Simulations}
\label{sec:3DMHD}

\subsection{Simulation Setup}

We describe the setup of our unstratified 3D MHD simulations, which are performed using the code {\it Athena++} \citep{Athena++}. The code solves the equations of ideal MHD with a fixed gravitational potential:
\begin{align}
\label{eq:cont}
\frac{\partial \rho}{\partial t} + \bfnabla \cdot (\rho \bfv) &= 0 \\
\label{eq:mom}
\frac{\partial (\rho \bfv)}{\partial t} + \bfnabla \cdot (\rho \bfv
\bfv) &= -\bfnabla \left( P + \frac{B^2}{2\mu} \right) + \frac{1}{\mu}
(\bfB \cdot \bfnabla) \bfB - \rho \bfnabla \Phi \\
\label{eq:induction}
\frac{\partial \bfB}{\partial t} &= \bfnabla \times (\bfv \times
\bfB).
\end{align}
We assume the equation of state is that of an isothermal ideal gas:
\begin{align}
\label{eq:state}
P &= \rho c_s^2.
\end{align}
Although this is a significant simplification, we are interested in BL dynamics and angular momentum transport, rather than the thermal structure of the BL. We dedimensionalize our simulation variables so that the radius of the star is at $R_* = 1$, and the Keplerian velocity at the surface of the star is $V_K(R_*) = 1$. We also choose the magnetic permeability of vacuum to be $\mu = 1$. 

We use a cylindrical coordinate system with logarithmic grid spacing in the radial direction. The boundary conditions are periodic in the $\phi$ and $z$ dimensions and ``do-nothing" in the $R$ dimension. The do-nothing boundary condition implies that hydrodynamic and MHD variables take on their initial values in the ghost zones for all time. The choice of do-nothing boundaries over reflecting or open boundaries is motivated by two considerations. First, do-nothing boundaries partially damp incident waves, in contrast to a perfectly reflecting boundary. This is important, because we do not want waves trapped between the boundary layer and the inner radial boundary to amplify by overreflection. Second, with an open boundary condition at the inner radius, we found that the star falls through the inner open boundary at an unacceptably fast rate. The reason for this is the difference between numerical and analytical hydrostatic equilibrium. Do-nothing boundaries are thus useful in our problem, because they damp incident waves and don't require precise initialization of numerical hydrostatic equilibrium. 

The domain of the simulation extends from $0.85 < R < 116$, $0 < \phi < 4 \pi /7$, $ -1/18 < z < 1/18$. The resolution in each dimension is $N_r \times N_\phi \times N_z = 4096 \times 768 \times 128$. The simulation is run for 936 Keplerian periods at the surface of the star, where $P_K (R_*) = 2 \pi$, in our dedimensionalized units. 

The initial state of our simulations consists of an accretion disk with a Keplerian rotation profile and a non-rotating star that are joined smoothly together:
\begin{align}
\Omega(R) = \begin{cases} 
      0 & R \leq 1- \Delta \\
      1 - (1-R)/\Delta & 1-\Delta \leq R \leq 1 \\
      R^{-3/2} & R\geq 1 
   \end{cases}.
\label{eq:rot}
\end{align}
Here, $R$ is the cylindrical radius and $\Delta=0.01$ is the {\it initial} width of the boundary layer. This is resolved with $\sim 8$ cells at the start of the simulation but widens as the simulation proceeds under the action of shear-acoustic instabilities. We inject the simulation with random perturbations to the steady-state density profile, which seeds the shear-acoustic instabilities in the BL and as well as the MRI instability in the disk.

The static gravitational potential is $\Phi(R) = -1/R$, and the effective potential is
\begin{align}
\Phi_\text{eff}(R) \equiv \Phi(R) - \int_{\infty}^R \Omega(R')^2 R' dR',
\end{align}  
where $\Omega(R)$ is the initial angular velocity profile (equation (\ref{eq:rot})). The initial hydrostatic equilibrium density profile is expressed in terms of the effective potential as
\begin{align}
\rho(R) = \exp\left(-\frac{\Phi_\text{eff}(R)}{c_s^2}\right),
\label{eq:rho}
\end{align}
where $c_s$ is the isothermal sound speed. The simulation is unstratified in the $z$-direction, because the gravitational potential we use is a function of the cylindrical radius only. Additionally, the initial density in the accretion disk is constant, since the initial rotation profile is exactly Keplerian there. We normalize the initial density in the disk to the value $\rho_\text{disk} = 1$.

We take $c_s = 0.1$ for the value of the isothermal sound speed in the 3D MHD simulation. This is potentially high by astrophysical standards (a white dwarf BL has $c_s \sim 0.02-0.05$). However, the radial pressure scale height,
\begin{align}
h_R \equiv \left|\frac{d \ln P}{dR}\right|^{-1}, 
\label{eq:scale_def}
\end{align}
takes the value
\begin{align}
h_{R,*} \sim R_*\left(\frac{c_s}{V_K(R_*)}\right)^2, 
\label{eq:scale}
\end{align}
inside the star.
With $c_s = 0.1$, the scale height in the star is $h_* \sim .01$, which is resolved with $\sim 8$ cells in the radial direction. Thus, the elevated value of the sound speed is necessary to resolve the scale height in the star. We also point out that there are $\sim 15$ scale heights between the outer edge of the star and the inner edge of the simulation domain. Thus, the density at the inner edge is $\rho_\text{max}/\rho_\text{disk} \sim 3 \times 10^7$, and the mass of the star within the simulation domain is much larger than the total mass accreted over the course of the simulation.

The initial magnetic field in the simulation is in the vertical direction and is given by 
\begin{align}
B_z(R) = \begin{cases} 
      0 & R \leq 1.25 \\
      (B_0/R) \zhat & 1.25 \leq R \leq 88 \\
      0 & R \geq 88 \\
   \end{cases}.
\label{eq:Binit}
\end{align}
The inner radius of the region of non-zero seed magnetic field is chosen to lie outside the star and the BL, so there is initially no magnetic field in these regions. However, field is advected into the BL as the simulation proceeds due to accretion induced by MRI turbulence in the disk. The outer radius of the region of non-zero seed magnetic field is chosen to lie well inside the outer radius of the simulation domain. Magnetic field does not diffuse to the outer boundary during the course of the simulation. Thus, there are no spurious effects that might arise due to boundary conditions. The magnetic field also never reaches the inner boundary of our simulation domain because accreted material forms a belt on the surface of the star and does not penetrate much below the BL. 

The fiducial seed magnetic field value in equation (\ref{eq:Binit}) is $B_0 = 2 \times 10^{-3}$. This means the magnetic $\beta$ parameter,
\begin{align}
\beta \equiv \frac{\rho c_s^2}{B^2/2\mu},
\end{align}
initially has the value $\beta \sim 5000$ in the inner parts of the accretion disk.  The vertical wavelength of the fastest growing axisymmetric MRI mode in the local approximation is given by
\begin{align}
\label{MRIlambda}
\lambda_{z,\text{MRI}} = \sqrt{\frac{16}{15}} \frac{2 \pi B}{\Omega
  \sqrt{\rho \mu}}
\end{align}
This is resolved with $\sim 16$ cells at the start of the simulation in the inner part of the disk, and $\lambda_{z,\text{MRI}}$ grows as $\propto R^{1/2}$ with radius.

\subsection{Verification}
\label{sec:verify}
Before analyzing our results in detail, we check to make sure the MRI in the disk and the acoustic modes excited in the BL are faithfully captured in our simulations. We begin by checking that the turbulent viscosity due to the MRI behaves as expected. The vertically-integrated Reynolds and magnetic stresses acting to drive accretion in the simulation can be expressed as
\begin{align}
\tau_{R \phi} \equiv \left \langle \rho v_R \left(v_\phi - R \Omega\right)\right \rangle -
\left \langle \frac{B_R B_\phi}{\mu} \right \rangle.
\label{eq:stress}
\end{align}
Here, $\Omega$ is the angular velocity, averaged over the $\phi$ and $z$ dimensions, and the brackets denote integration over the $z$-dimension and averaging over the $\phi$-dimension. We shall also find it useful to split $\tau_{R \phi}$ into purely hydrodynamical (subscript H) and purely magnetic (subscript B) components:
\begin{align}
\label{eq:Hstress}
\tau_{R \phi,H} &\equiv \left \langle \rho v_R \left(v_\phi - R \Omega\right)\right \rangle \\
\label{eq:Bstress}
\tau_{R \phi,B} &\equiv -\left \langle\frac{B_R B_\phi}{\mu} \right \rangle.
\end{align}
The total $R\phi$ stress is simply the sum of the individual components:
 \begin{align}
 \tau_{R \phi} =  \tau_{R \phi,H} +  \tau_{R \phi,B}.
 \end{align}
 
The one-dimensional equation describing angular momentum transport is
\begin{align}
\frac{\partial}{\partial t}\left(R^2 \Sigma \Omega \right)
&= -\frac{1}{R} \frac{\partial}{\partial R} \left(R^3 \Omega \Sigma v_R + R^2  \tau_{R\phi} \right),
\label{1Dangmom}
\end{align}
where $\tau_{R\phi}$ due to turbulent stresses is given by equation (\ref{eq:stress}) \citep{BalbusPapaloizou}. Here, $\Sigma = \langle \rho \rangle$ is the 1D disk surface density, and $v_R = \Sigma^{-1} \langle \rho v_R \rangle$ is the 1D density-weighted radial velocity. Equation (\ref{1Dangmom}) has the same form as the 1D equation of viscous accretion disk theory, except that the stress in viscous theory is given by
\begin{align}
\tau_{R \phi} = -\nu \Sigma R \frac{d \Omega}{d R}.
\label{eq:tau_disk}
\end{align} 
The viscosity in the accretion disk is typically parameterized by
\begin{align}
\nu = \alpha c_s H,
\label{eq:nu_disk}
\end{align}
where $H$ is the vertical scale height in the disk and $\alpha < 1$ is a dimensionless constant. 

\begin{figure}[!t]
\centering
\subfigure{\includegraphics[width=0.49\textwidth]{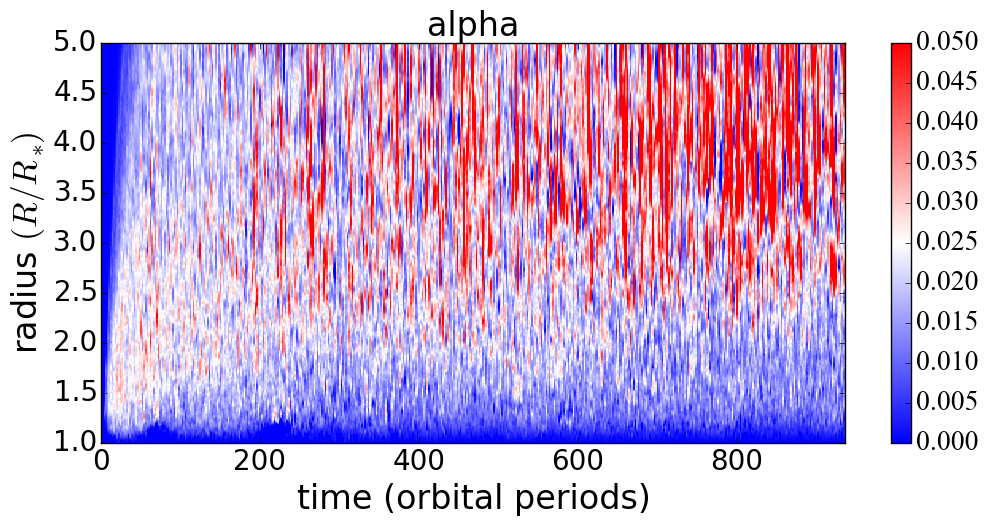}}
\subfigure{\includegraphics[width=0.49\textwidth]{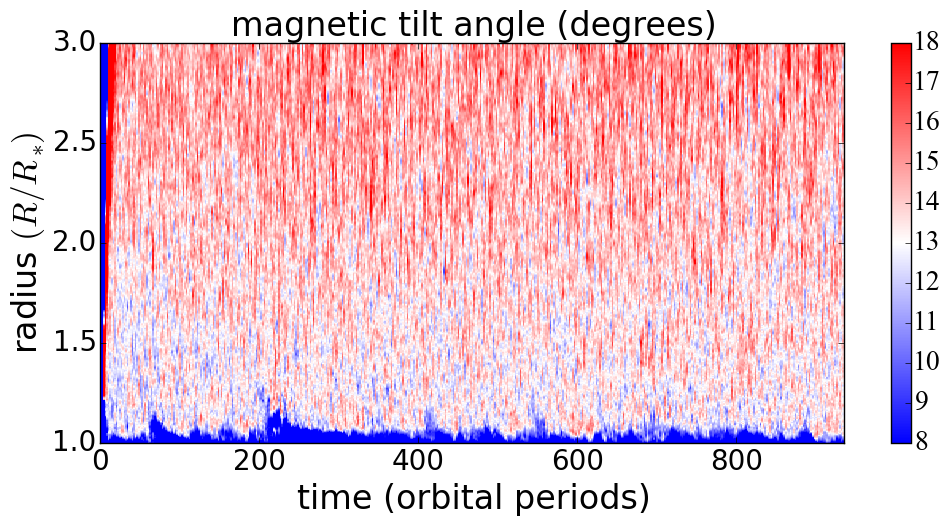}}
\caption{Left panel: spacetime plot of the effective value of alpha in the inner part of the disk. Right panel: spacetime plot of the magnetic pitch angle in the inner part of the disk.}
\label{fig:MRI}
\end{figure}

If we measure the stresses according to equation (\ref{eq:stress}) within the simulation, we can compute the effective value of $\nu$ in the accretion disk by using equation (\ref{eq:tau_disk}). We can then calculate the effective value of $\alpha$ by using equation (\ref{eq:nu_disk}) and setting $H$ to be the vertical extent of the simulation domain. The left panel of Fig.\ \ref{fig:MRI} shows a spacetime plot of $\alpha$ in the disk ($R>1$), averaged over the $\phi$ and $z$ dimensions. The $x$-axis is $t/P_{K,*}$, and the $y$-axis is the cylindrical radius. After MRI turbulence develops, we have $\alpha \sim .01-.05$ in the accretion disk. 

A different check that determines the convergence of the MRI in the saturated state is the magnetic pitch angle. \cite{GuanGammie} define the pitch angle as
\begin{align}
\theta_B \equiv \frac{\sin^{-1}\left(\alpha_B \beta\right)}{2},
\end{align}
where $\alpha_B$ is calculated using {\it only} the magnetic component of the stress tensor (equation (\ref{eq:Bstress})) together with the definition of $\alpha$ in terms of the stress via equations (\ref{eq:tau_disk}) and (\ref{eq:nu_disk}). The magnetic pitch angle is a good indicator of convergence for the MRI, and for converged, unstratified shearing box simulations $\theta_B \approx 14-16 ^{\circ}$. The right panel of Fig.\ \ref{fig:MRI} shows a spacetime plot of the magnetic pitch angle in our unstratified simulation averaged over the $\phi$ and $z$ dimensions. The $x$-axis shows the time in orbital periods, and the $y$-axis shows the radius. The pitch angle in the innermost parts of the disk is close to the resolved shearing box value.

\begin{figure}[!t]
\centering
\includegraphics[width=0.49\textwidth]{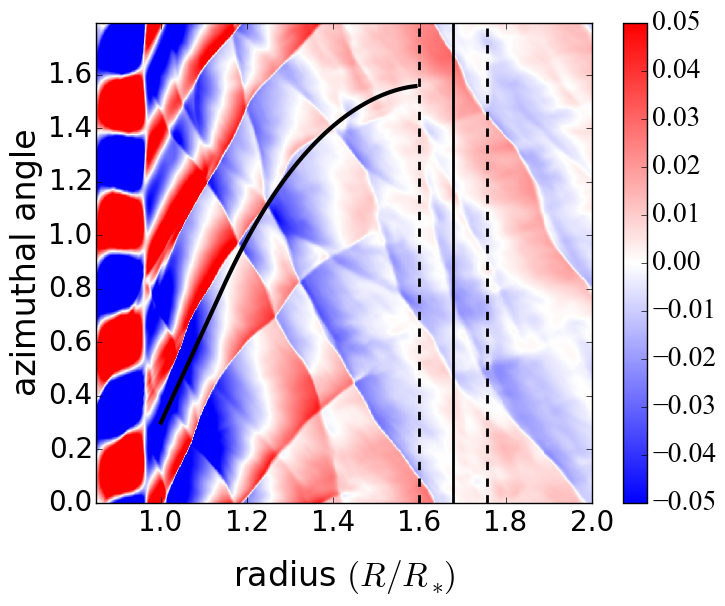}
\caption{Color shows $v_R \sqrt{\rho}$ in the disk averaged over the $z$-dimensions with $R$ on the $x$-axis and $\phi$ on the $y$-axis  at $t=204$ orbits in the MHD simulation. An $m=14$ mode is apparent in the star (limits in azimuthal direction are $0 < \phi < 4 \pi/7$). The vertical lines show the corotation radius (solid) and Lindblad radii (dashed) in the accretion disk for $\Omega_P = 0.46$. The solid black curve shows a sample wavefront that has been calculated using the WKB approximation for acoustic wave propagation in the disk.}
\label{mode_fig}
\end{figure}

Next we consider shear-acoustic modes excited in the BL. Fig.\ \ref{mode_fig} shows an image of $v_R \sqrt{\rho} $ averaged over the $z$-dimension at $t=204P_{K,*}$ in the unstratified MHD simulation\footnote{Note that $v_R \sqrt{\rho} $ is a good quantity to measure, because it is roughly constant due to conservation of energy flux between $R=.85$ (inner boundary) and $R\approx1$ (surface of the star), even though the density increases by orders of magnitude over this region.}. A shear-acoustic mode with azimuthal pattern number $m=14$ is visible. This mode is sourced in the BL and has a pattern speed of $\Omega_P = 0.46$ as measured in the simulation. The measured pattern speed of $\Omega_P = 0.46$ agrees well with the predicted pattern speed for an $m=14$ mode given by equation (38) of \cite{BRS1}. The solid vertical lines indicate the mode's corotation radius and the dashed vertical lines indicate its Lindblad radii in the accretion disk. The mode also has a corotation radius in the boundary layer (not shown). 

Another interesting question concerns the dominance of the $m=14$ mode in the simulation. Although a spectrum of shear-acoustic modes is expected to be excited in the BL, \cite{acousticCFS} showed that a sound wave which reflects off the inner Lindblad radius in the disk and returns in phase with the outgoing mode in the BL can be pumped to higher amplitudes in the disk with same forcing amplitude in the BL. The criterion for this to occur is
\begin{align}
\Delta \phi \approx 2\pi n/ m,
\label{eq:wrap_crit}
\end{align}
where $\Delta \phi$ is the angle traversed by the sound wave in the disk in the azimuthal direction as it travels from the BL to the inner Lindblad radius and back. The integer $n$ determines the number of radial nodes in the disk as the acoustic mode wraps back on itself. Modes that are close to satisfying the criterion of equation (\ref{eq:wrap_crit}) are pumped to higher amplitudes in the disk, which appears to be the case for the $m=14$ mode in Fig.\ \ref{mode_fig}.

\section{Belt of High Angular Momentum Material}
\label{sec:belt}

\subsection{Magnetic Fields}

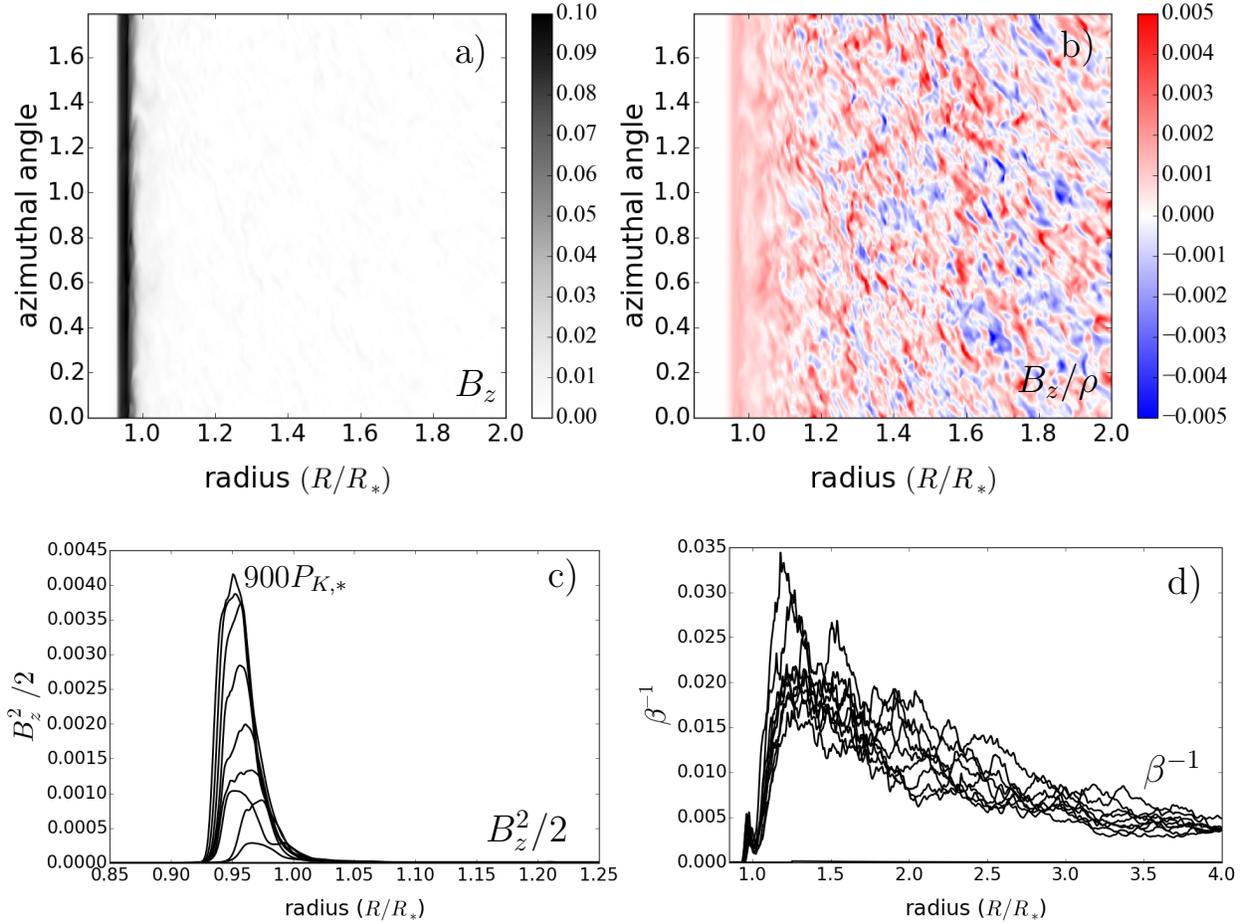
\begin{figure}[!t]
\centering
\subfigure{\begin{overpic}
	[width=0.48\textwidth]{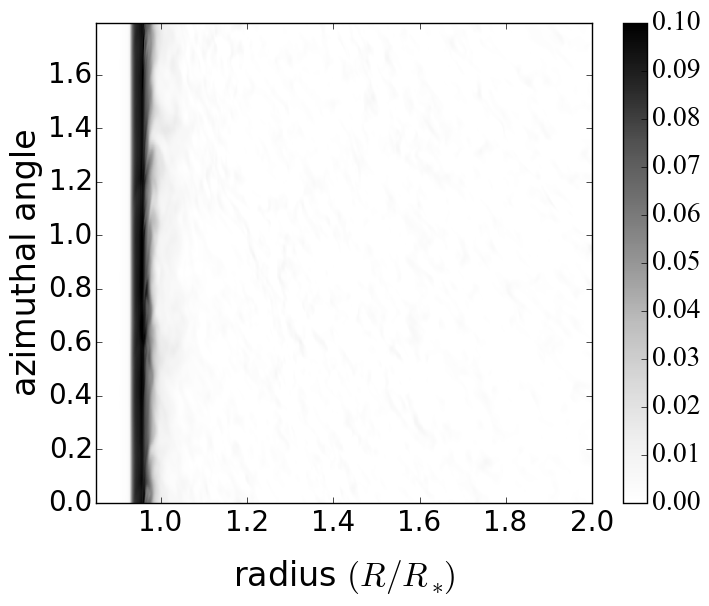}
	\put(75,74){\Large a)}
	\put(75,18){\Large $B_z$}
\end{overpic}}
\subfigure{\begin{overpic}
	[width=0.5\textwidth]{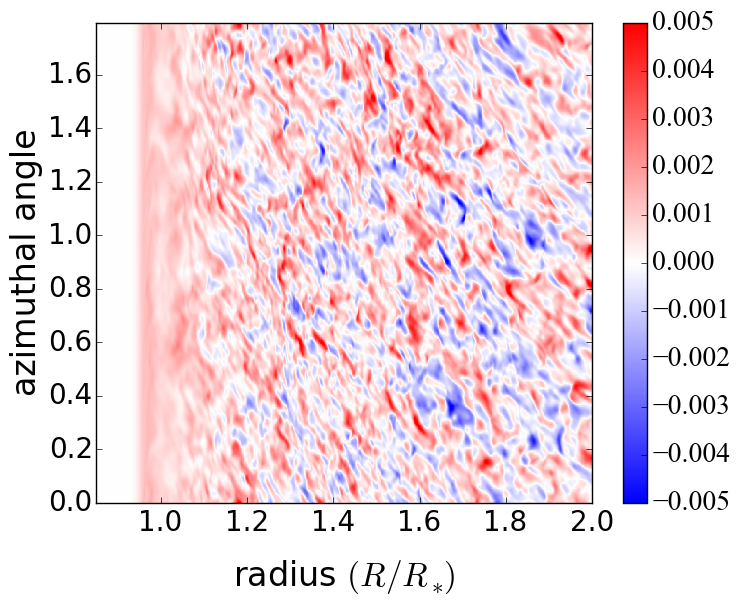}
	\put(72,72){\Large b)}
	\put(65,18){\Large $B_z/\rho$}
\end{overpic}}
\subfigure{\begin{overpic}
	[width=0.5\textwidth]{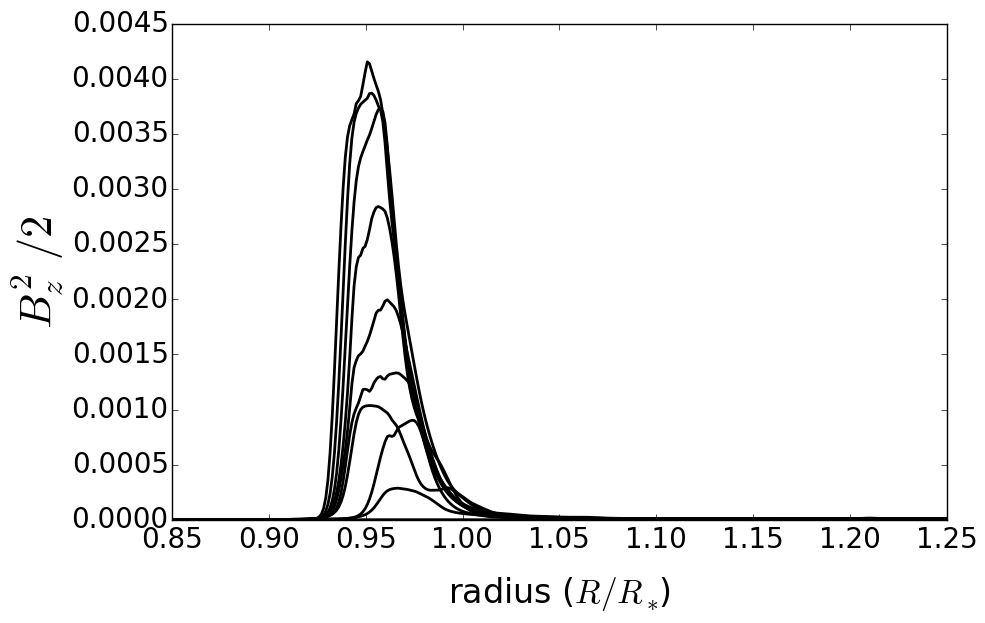}
	\put(88,55){\Large c)}
	\put(78,14){\Large $B_z^2/2$}
	\put(39,55){\large $900 P_{K,*}$}
\end{overpic}}
\subfigure{\begin{overpic}
	[width=0.49\textwidth]{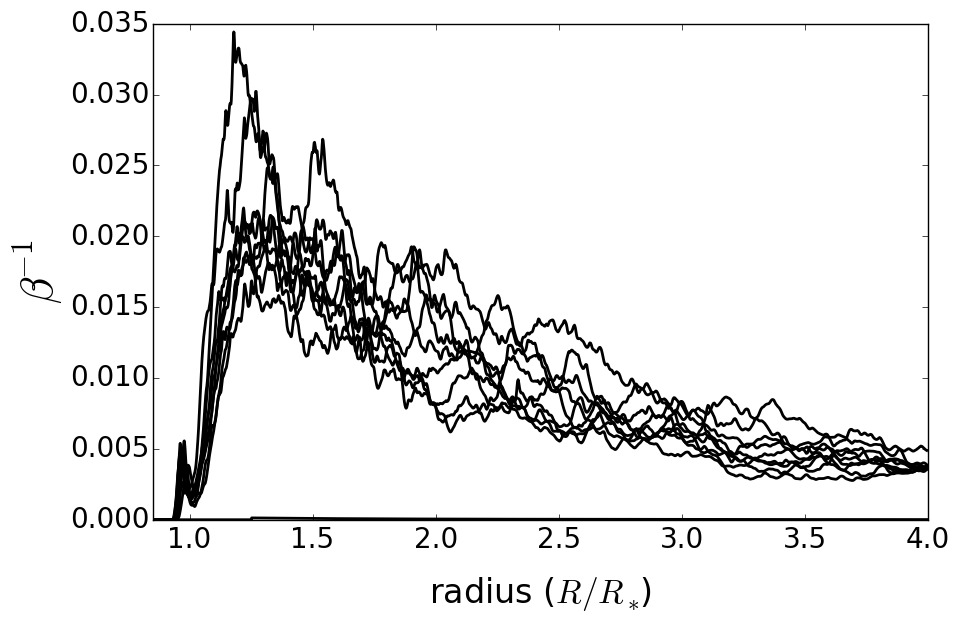}
	\put(88,55){\Large d)}
	\put(84,24){\Large $\beta^{-1}$}
\end{overpic}}
\caption{a) $B_z$ averaged over $z$ at $t=930P_{K,*}$ in the 3D MHD simulation. The belt of accreted material in the BL is visible as the black band inside of $R = 1$. b) $B_z/\rho$ averaged over $z$ at $t=930P_{K,*}$ in the 3D MHD simulation. c) Curves show the magnetic energy density ($B_z^2/2$) averaged over $\phi$ and $z$ at between $t=0$ and $t=900 P_{K,*}$ in intervals of $100 P_{K,*}$. The highest amplitude labeled curve corresponds to $t=900 P_{K,*}$. d) Curves show the ratio of the magnetic energy density to the thermal energy density ($\beta^{-1}$) averaged over $\phi$ and $z$ between $t=0$ and $t=900 P_{K,*}$ in intervals of $100 P_{K,*}$. The value of $\beta^{-1}$ has a small bump in the boundary layer around $R=1$. However, this is still small compared to its value in the inner disk.}
\label{fig:flux}
\end{figure}

As the simulation proceeds, material from the disk is accreted onto the surface of the star. Because the disk initially has a net vertical flux that is frozen into the fluid, the flux is dragged along together with the fluid material. Thus, one way to visualize the accreted material is with images of the vertical component of the magnetic field, $B_z$. This is shown in panel a of Fig.\ \ref{fig:flux} at $t=900 P_{K,*}$ (towards the end of the simulation). Because the magnetic flux is frozen into the fluid, the vertically-integrated value of $B_z/\rho$ is conserved even as material loses angular momentum and is compressed in the BL. The amplification of $B_z$ in the BL and inner disk that is apparent in the image of $B_z$ in panel a of Fig.\ \ref{fig:flux} is not seen in the image of $B_z/\rho$ in panel b. The increase in $B_z$ in the BL is explained by conservation of magnetic flux dragged into the BL and is not due to magnetic field amplification. The amplitude of $B_z$ traces the accreted material from the disk. 

Another important point is that the ratio of the magnetic energy to the thermal energy in the BL remains small, despite accumulation of $B_z$ in the BL. Panels c and d of Fig.\ \ref{fig:flux} show plots of $B_z^2/2$ and $\beta^{-1}$, respectively, between $t=0$ and $t=900 P_{K,*}$ in intervals of $100 P_{K_*}$ (each curve corresponds to a particular snapshot in time e.g.\ $t=0$, $t=100 P_{K,*}$, ..., $t=900 P_{K,*}$). In panel c, the peak value of $B_z^2/2$ in the BL (around $R=1$) increases monotonically for the duration of the simulation. On the other hand, this behavior is not observed in panel d for $\beta^{-1}$, which peaks in the inner disk and shows only a small bump in the BL around $R=1$. In addition, the amplitude variations of $\beta^{-1}$ once MRI turbulence has developed are only a factor of a few, and a monotonic trend in time is not observed.

\begin{figure}[!t]
\centering
\subfigure{\begin{overpic}
	[width=0.5\textwidth]{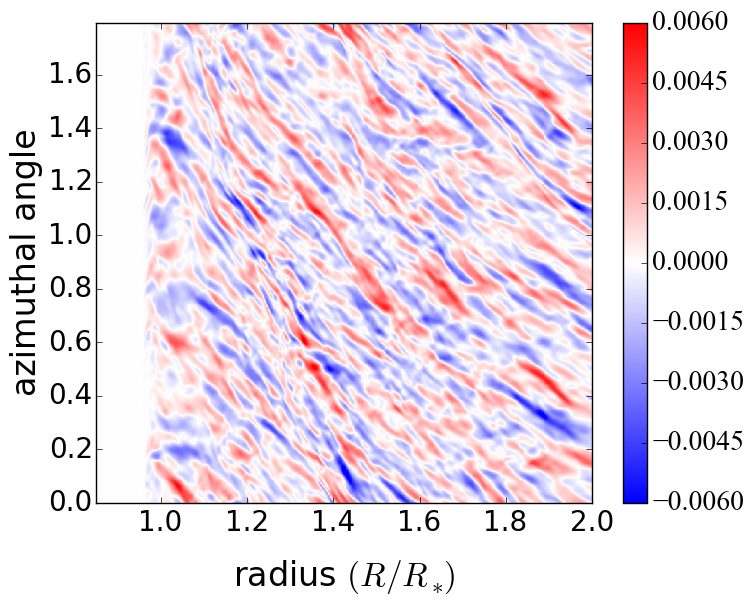}
	\put(68,70){\Large $B_R$}
	\put(15,70){\Large a)}
\end{overpic}}
\subfigure{\begin{overpic}
	[width=0.49\textwidth]{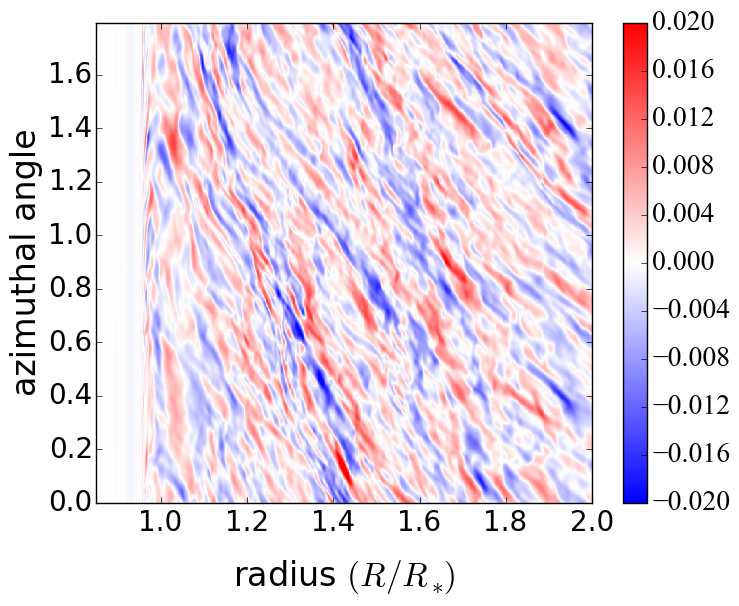}
	\put(70,70){\Large $B_\phi$}
	\put(15,70){\Large b)}
\end{overpic}}
\subfigure{\begin{overpic}
	[width=0.49\textwidth]{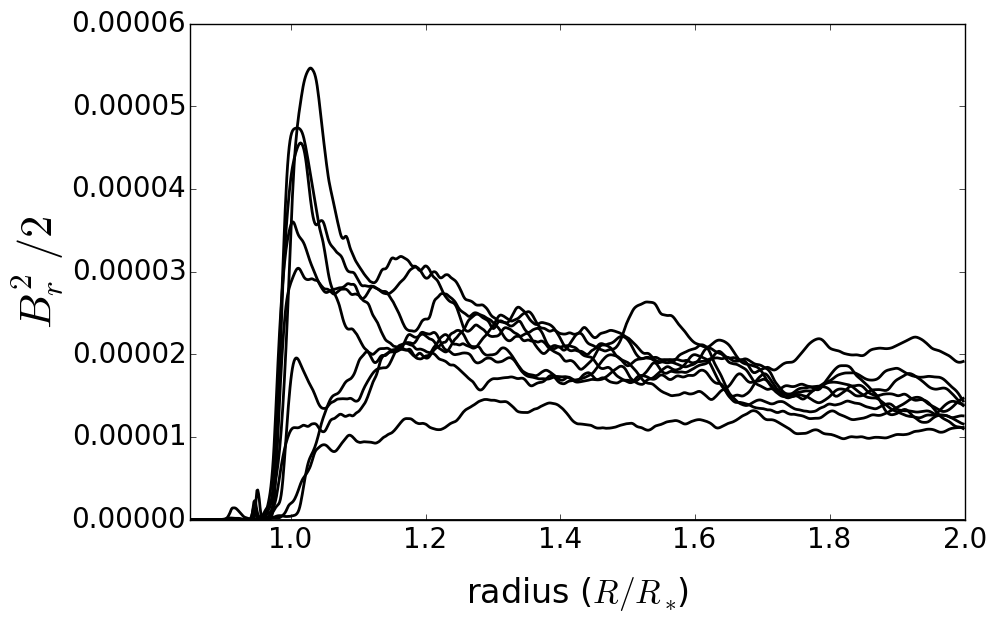}
	\put(80,53){\Large $B_R^2/2$}
	\put(20,54){\Large c)}
\end{overpic}}
\subfigure{\begin{overpic}
	[width=0.49\textwidth]{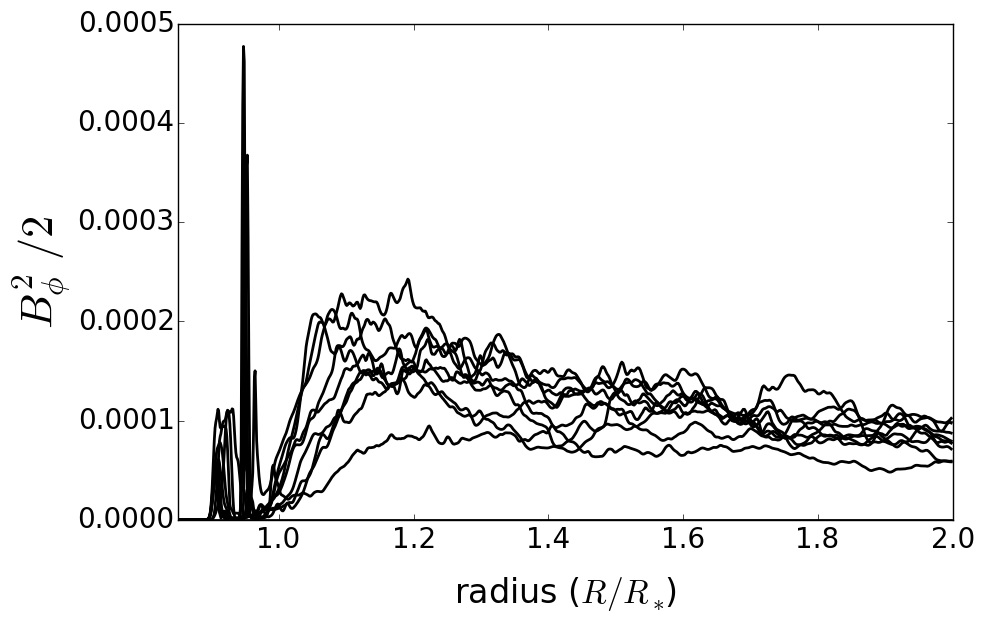}
	\put(80,53){\Large $B_\phi^2/2$}
	\put(19,54){\Large d)}
\end{overpic}}
\caption{a) $B_R$ averaged over $z$ at $t=930P_{K,*}$ in the 3D MHD simulation. b) $B_\phi$ averaged over $z$ at $t=930P_{K,*}$ in the 3D MHD simulation. c) Curves show $B_R^2/2$ ($R$-component of the magnetic energy) averaged over $\phi$ and $z$ between $t=0$ and $t=900 P_{K,*}$ in intervals of $100 P_{K,*}$. d) Same as panel c, but for $B_\phi^2/2$ ($\phi$-component of the magnetic energy).}
\label{fig:flux2}
\end{figure}

The plot of $\beta^{-1}$ in panel d of Fig.\ \ref{fig:flux} suggests that magnetic fields are not amplified in the BL within our simulation. To further demonstrate this point, we show that neither $B_R$ nor $B_\phi$ undergo significant amplification in the BL. Panels a and b of Fig.\ \ref{fig:flux2} show images of $B_R$ and $B_\phi$ averaged over $z$ at $t=936P_{K,*}$. The magnetic field in both cases transitions smoothly from its disk value set by MRI turbulence to zero in the star. Panels c and d of Fig.\ \ref{fig:flux2} show plots of $B_R^2/2$ and $B_\phi^2/2$ (the $R$ and $\phi$ components of the magnetic energy) averaged over $\phi$ and $z$ between $t=0$ and $t=900 P_{K,*}$ in intervals of $100 P_{K_*}$. Although panel d shows transient spikes in the amplitude of $B_\phi^2/2$ within the BL, sustained amplification of $B_R^2$ or $B_\phi^2$ is not observed in the simulation.

It is evident from the image of $B_z$ in Fig.\ \ref{fig:flux} that the accreted material forms a belt on the surface of the star. As we now demonstrate, the material in this belt does not efficiently give up its angular momentum to the star, despite the large shear present in the BL and the shear-acoustic instabilities excited there.

\subsection{Flow of Angular Momentum}

To understand the flow of angular momentum in the simulation, it is useful to define the stress angular momentum current,
\begin{align}
C_L \equiv 2 \pi R^2 \tau_{r \phi},
\end{align}
where $\tau_{r \phi}$ is defined via equation (\ref{eq:stress}). Note that $C_L$ is just the angular momentum per unit time transferred by the stress $\tau_{R\phi}$. The stress angular momentum current is constant for waves in the $R-\phi$ plane in the absence of damping or amplification. We also define the hydrodynamical and magnetic components of the stress angular momentum current as 
\begin{align}
\label{eq:CLH}
C_{L,H} &\equiv 2 \pi R^2 \tau_{r \phi,H} \\
\label{eq:CLB}
C_{L,B} &\equiv 2 \pi R^2 \tau_{r \phi,B},
\end{align}
where $\tau_{r \phi,H}$ and $\tau_{r \phi,B}$ are defined in equations (\ref{eq:Hstress}) and (\ref{eq:Bstress}), respectively. Because our 3D MHD simulation is unstratified in the vertical direction, we display the values of $C_L$, $C_{L,H}$ and $C_{L,B}$ per unit $z$, so they are independent of the box height. That is we define $C_L \rightarrow C_L/\Delta z$ with $\Delta z = 1/9$.

\begin{figure}[!t]
\centering
\subfigure{\begin{overpic}
	[width=0.98\textwidth]{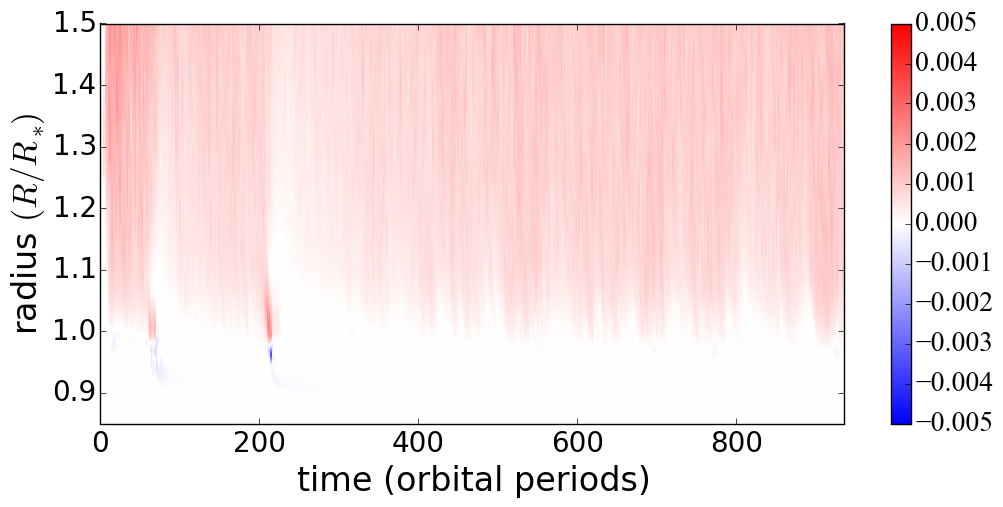}
	\put(73,42){\huge $C_{L,B}$}
\end{overpic}}
\subfigure{\begin{overpic}
	[width=0.98\textwidth]{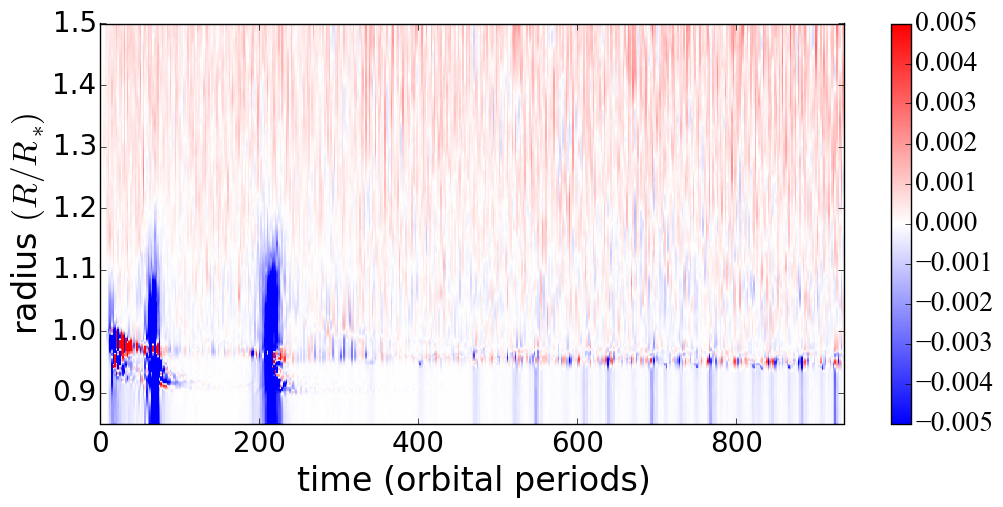}
	\put(73,42){\huge $C_{L,H}$}
\end{overpic}}
\caption{Upper panel: spacetime plot of magnetic component of the stress angular momentum current (equation \ref{eq:CLB}), which quantifies the magnetic component of the angular momentum transport. Lower panel: spacetime plot of the hydrodynamical component of the stress angular momentum current (equation \ref{eq:CLH}) which quantifies angular momentum transport by hydrodynamical Reynolds stresses and waves. Outbursts of shear-acoustic instability observed at $t=220$ and $t=70$ result in increased Reynolds stress in the BL and inner disk.}
\label{fig:stress}
\end{figure}

Fig.\ \ref{fig:stress} shows spacetime plots of $C_{L,B}$ (upper panel) and $C_{L,H}$ (lower panel) for the 3D MHD simulation. From the upper panel, we see that $C_{L,B}$ is positive and is roughly constant in time within the disk ($R \gtrsim 1$). This is expected for steady state MRI turbulence, which transports angular momentum outward. Inside the star ($R \lesssim 1$) $C_{L,B}$ vanishes, since the magnetic field is zero there. Inside the BL ($R \sim 1$), we see no evidence for a significant magnetic component to the stress, and $C_{L,B}$ transitions smoothly from its disk value to zero inside in the star. Thus, despite advection of vertical field into the BL (Fig.\ \ref{fig:flux}), there is little transport of angular momentum by magnetic stresses there.

In contrast to $C_{L,B}$, the spacetime plot of $C_{L,H}$ is significantly more complicated. The bands of blue inside the star ($R \lesssim 1$) in the plot of $C_{L,B}$, correspond to a hydrodynamical angular momentum current carried by gravitosonic waves (sound waves modified by radial stratification) excited via the shear-acoustic mechanism in the BL. The fact that $C_{L,H} < 0$ in the star means the waves carry positive angular momentum inward (i.e.\ in the negative radial direction), and act to spin up the star. On the other hand, acoustic waves excited in the BL that propagate into the disk\footnote{The acoustic waves are spiral density waves without self-gravity.} have $\Omega_P < \Omega_K(R_*)$ inside of the inner Lindblad radius where they reflect. Thus, they transport negative angular momentum outward (i.e.\ in the positive radial direction), and $C_{L,H}$ due to the waves in the inner part of the disk is also negative.

One of the most striking differences between angular momentum transport due to acoustic waves excited in the BL versus MRI turbulence is how time-variable the former is compared to the latter. From the image of $C_{L,H}$ in Fig.\ \ref{fig:stress}, we see two major ``outbursts" of shear-acoustic instability in the BL at $t \approx 70$ and $t \approx 220$. During these outbursts, waves are excited to high amplitudes and transport angular momentum at a faster rate than MRI in the disk. However after the second large outburst, there is a ``dry spell" between $t=250-500$ when angular momentum transport via waves in the star is at a much lower rate than via MRI turbulence in the disk. Subsequently, between $t=500$ and the end of the simulation, we see many ``mini-outbursts" separated in time by ``mini-dry spells".

\begin{figure}[!t]
\centering
\subfigure{\begin{overpic}
	[width=0.58\textwidth]{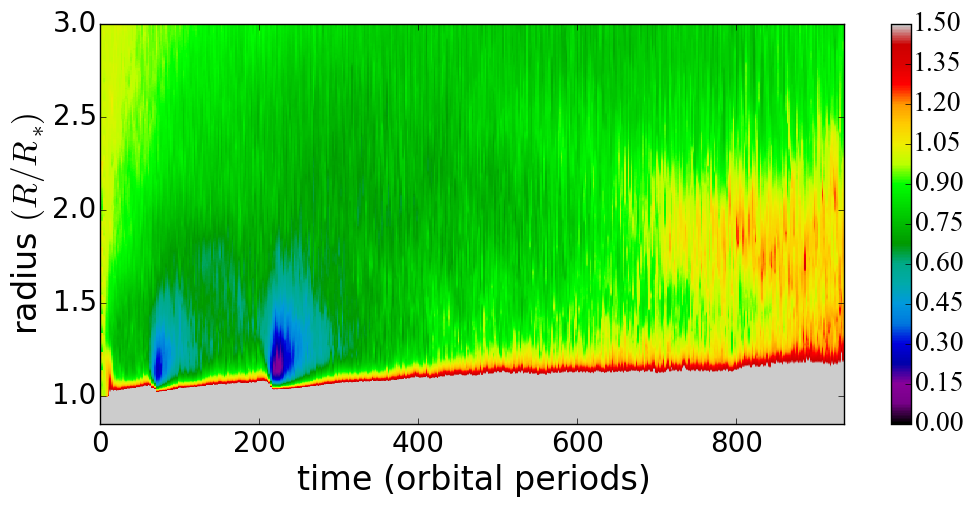}
	\put(80,43){\Large $\Sigma$}
\end{overpic}}
\subfigure{\begin{overpic}
	[width=0.4\textwidth]{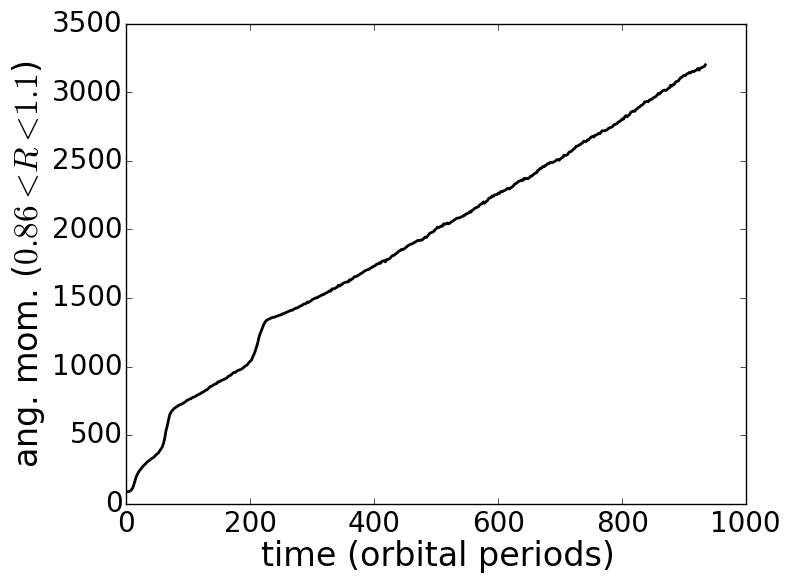}
\end{overpic}}
\caption{Left panel: Spacetime plot of the surface density in the disk and the inner part of the BL in the 3D MHD simulation. The gray high density region at smaller radii is the star and the region with $\Sigma \sim 1$ at larger radii is the accretion disk. The outbursts of shear-acoustic instability at $t=70$ and $t=220$ deplete the inner disk of mass. Right panel: Total angular momentum between $.86 < R < 1.1$ as a function of time in the 3D MHD simulation. This is a proxy for the total angular momentum accreted from the disk onto the surface of the star. The increase in angular momentum is approximately linear in time, except for two sharp jumps. The linear trend is due to accretion of through the disk under the action of MRI turbulence. The jumps are coincident with outbursts of shear-acoustic instability at $t=70$ and $t=220$. }
\label{dens_st_fig}
\end{figure} 

The left panel Fig.\ \ref{dens_st_fig} shows a spacetime plot of the surface density in the inner part of the simulation domain. During episodes of wave driven accretion when hydrodynamical stresses are high (see Fig.\ \ref{fig:stress}) the density in the innermost parts of the disk is depleted. This can be seen as bluish-purple regions in Fig.\ \ref{dens_st_fig} around $R \gtrsim 1$ at $t = 70$ and $t = 220$. After each outburst of shear-acoustic instability, the density rises slowly as the depleted regions are filled in by material accreted from the outer part of the disk due to MRI turbulence.

The right panel of Fig.\ \ref{dens_st_fig} shows a plot of the total angular momentum accreted onto the surface of the star. During outbursts of shear-acoustic instability at $t=70$ and $t=220$, there is a rapid rise in the accreted angular momentum. This demonstrates that shear-acoustic instabilities are extremely efficient at angular momentum transport in the inner part of the disk, but only when waves in the BL are excited to high amplitude. Outside of the outbursts, accreted angular momentum increases approximately linearly in time. This indicates MRI turbulence in the disk acts like an effective viscosity and leads to an approximately constant mass accretion rate onto the surface of the star. 

\subsection{Belt Formation}

We cannot exclude the possibility of more large outbursts of shear-acoustic instability on timescales that are long compared to the duration of our simulation. Nevertheless, we may ask whether the 3D MHD simulation reaches a quasi-steady state after the outbursts have stopped (i.e. for $t \gtrsim 500 P_{K,*}$). In particular, do waves excited in the BL during the ``mini-outburst" and ``mini-dry spell" phase transport angular momentum through the BL into the star at the same rate that it is accreted onto the BL from the disk?

To provide an answer to this question, we average $C_{L}$ between $R=.86-.92$ in radius and between $t=600$ and the end of the simulation in time, which yields $C_{L} \approx -3 \times 10^{-4}$. This is the angular momentum transport rate in the star. Computing the angular momentum transport rate through the disk in the MHD simulation is trickier, because both advected and viscous stresses are important. These two components of the stress have comparable magnitude and opposite sign, making it tricky to accurately calculate their sum, because they nearly cancel one another. However, we can use viscous disk theory (\S \ref{sec:viscAMT}) to assess whether or not the system is in a steady state. In particular, using equations (\ref{Mdisk}) and (\ref{Jdiskapprox}), assuming $\alpha \sim .03$ in the disk as suggested by Fig.\ \ref{fig:MRI}, and taking $\rho_\text{disk}=1$, the rate of accumulation of angular momentum in the BL (per unit $z$) is $\dot{J} \approx 3 \times 10^{-3} (\alpha/.03)(\rho_\text{disk}/1)$. This is an order of magnitude larger than the rate at which waves excited in the BL carry angular momentum into the star. 

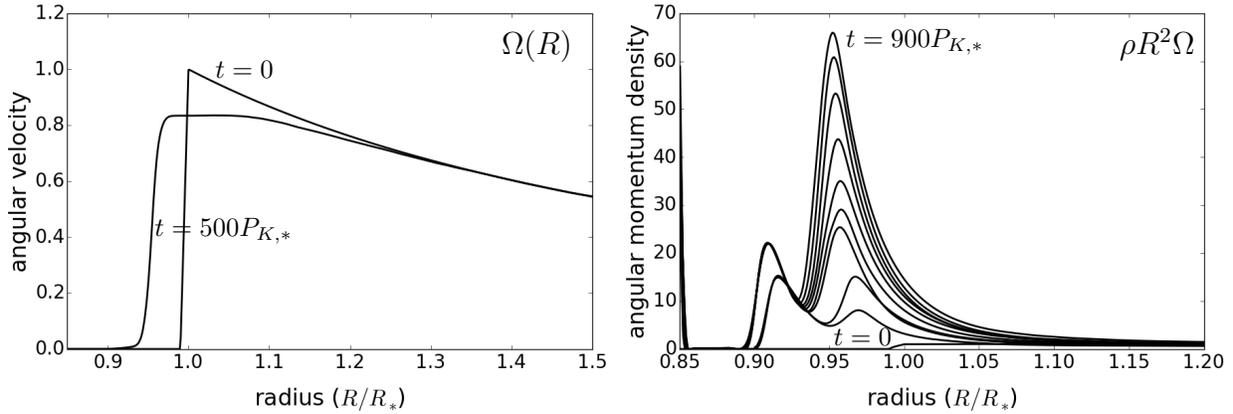
\begin{figure}[!t]
\centering
\subfigure{\begin{overpic}
	[width=0.49\textwidth]{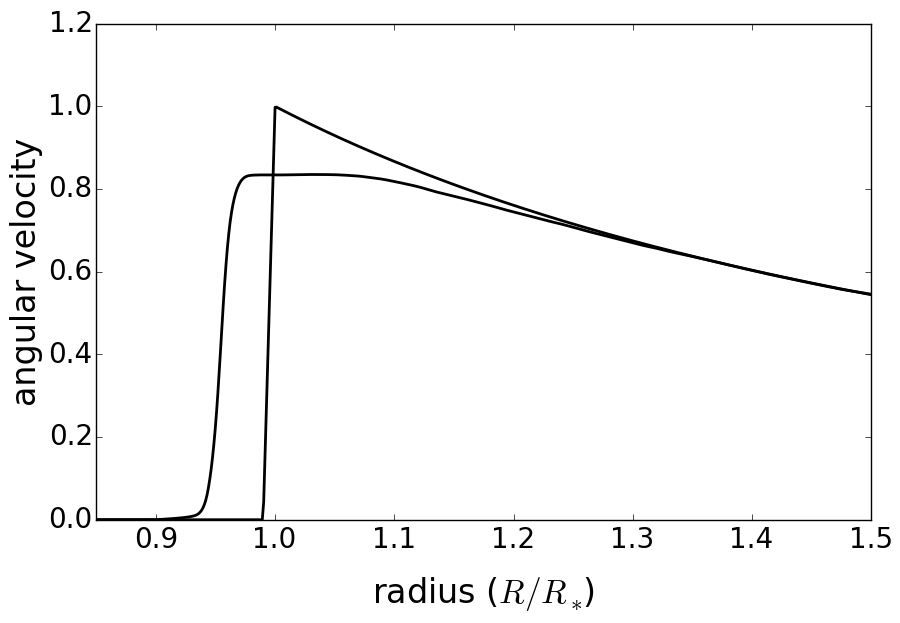}
	\put(35,56){\small $t=0$}
	\put(25,30){\small $t=500 P_{K,*}$}
	\put(82,60){\large $\Omega(R)$}
\end{overpic}}
\subfigure{\begin{overpic}
	[width=0.49\textwidth]{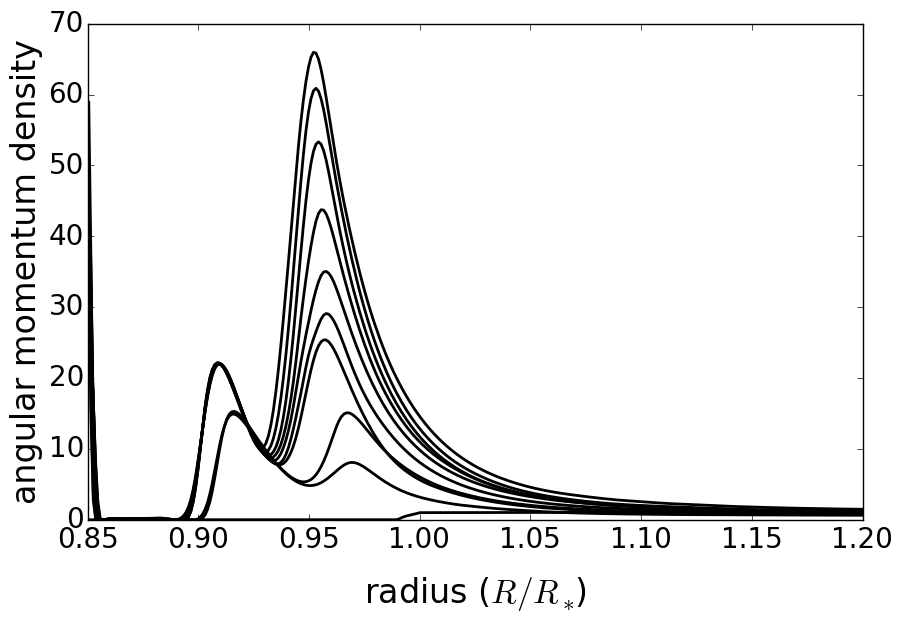}
	\put(82,60){\large $\rho R^2 \Omega$}
	\put(37,61){\small $t=900 P_{K,*}$}
	\put(35,12){\small $t=0$}
\end{overpic}}
\caption{Left panel: angular velocity at the start of the 3D MHD simulation ($t=0$) and at 500 orbital periods ($t=500 P_{K,*}$). At $t=0$ the angular velocity is Keplerian for $R > 1$ and $\Omega = 0$ for $R < .99$. At $t=500P_{K,*}$ the angular velocity is significantly sub-Keplerian in the inner part of the disk but is essentially Keplerian beyond $R \gtrsim 1.2$. Right panel: angular momentum density ($\rho R^2 \Omega$) in intervals of 100 orbital periods from $t=0$ to $t=900 P_{K,*}$. There is a belt of angular momentum in the BL that grows monotonically in time during the simulation.}
\label{fig:rot}
\end{figure}

The mismatch in the rate at which angular momentum enters the BL and the rate at which it is transported into the star leads to accumulation of angular momentum in the BL. The right panel of Fig.\ \ref{fig:rot} shows plots of the angular momentum profile in the 3D MHD simulation at intervals of $t=100P_{K,*}$. As the simulation proceeds, a belt of rapidly rotating material develops on the surface of the star and grows monotonically in time. In addition to the main belt of accreted angular momentum between $.94 \lesssim R \lesssim 1.05$, there is a bump in the angular momentum density at $.9 \lesssim R \lesssim .94$. The bump is formed during the first outburst of shear-acoustic instability at $t \approx 70$ and then is enhanced and moves inward during the second outburst at $t \approx 220$. 

For comparison, the left panel of Fig.\ \ref{fig:rot} shows the angular velocity at the start of the simulation and at $t=500P_{K,*}$. As the simulation proceeds, a plateau of constant $\Omega$ develops in the angular velocity profile in the innermost part of the disk ($.95 \lesssim R \lesssim 1.1$). This plateau adjoins the outer radial edge of the BL, and material in this region is part of the main angular momentum belt. The bump in the angular momentum profile (right panel of Fig.\ \ref{fig:rot}) is adjacent to the lower edge of the BL between $.9 \lesssim R \lesssim .95$. The fact that the angular velocity is constant in the plateau region spanning the angular momentum belt suggests there is a physical process enforcing corotation within the belt. Otherwise, one would expect specific angular momentum to be conserved, not $\Omega$. This is potentially related to the non-amplification of magnetic field in the BL, which is an interesting topic for future exploration.

\section{2D Viscous Simulations}
\label{sec:visc}

\subsection{Simulation Setup}
\label{sec:viscAMT}
In order to better understand the results of the 3D MHD simulation, we perform 2D viscous hydro runs using cylindrical coordinates in the $R-\phi$ plane. The 2D viscous simulations contain no magnetic fields. Instead, accretion and angular momentum transport in the disk are facilitated via a viscous stress. This has the advantage that we can control both the mass transport and the angular momentum transport rates through the disk. As a result, we can accurately determine the fraction of the angular momentum current carried by the waves in the BL and the star, which is tricky in the 3D MHD simulation due to the difficulty in computing the angular momentum transport rate through the disk. 

The 2D momentum equation including viscosity can be written in vector form as
\begin{align}
\frac{\partial (\Sigma \bfv)}{\partial t} + \bfnabla \cdot (\Sigma \bfv
\bfv) &= -\bfnabla P - \bfnabla \cdot \boldsymbol{\tau}.
\label{eq:momvis}
\end{align}
Here, $\Sigma$ is the 2D surface density, $P$ is the pressure integrated over $z$, and $\boldsymbol{\tau}$ is the viscous stress tensor\footnote{We write the viscous stress tensor with a negative sign compared to the usual formulation to ensure consistency with the definition of the turbulent stress in equation (\ref{eq:stress}). It also has the intuitive feature that a positive $R\phi$-stress means an outward viscous transport of angular momentum.}.  

We approximate viscous angular momentum transport as a 1D process by averaging equation (\ref{eq:momvis}) over the $\phi$-dimension at each timestep in the 2D viscous hydro simulations. In this case, only the $RR$ and $R\phi$-components of the viscous stress contribute to the momentum equation:
\begin{align}
\tau_{R\phi}(R)  &\equiv -\nu \left \langle \Sigma R \frac{\partial \Omega}{\partial R} \right \rangle, \\ 
\tau_{RR}(R) &\equiv -2 \nu \left \langle \Sigma \left(\frac{\partial v_R}{\partial R} -\frac{1}{3}\bfnabla \cdot \bfv \right) \right \rangle.
\end{align}
The viscosity parameter in our simulations takes different constant values depending on whether the viscous stress $\tau_{R\phi} < 0$ or $\tau_{R\phi} > 0$: 
\begin{align}
\nu(R) = \begin{cases} 
      \nu_\text{disk}, & \tau_{R\phi}(R) > 0 \\
      \nu_\text{BL}, & \tau_{R\phi}(R) < 0
   \end{cases}.
\end{align} 
When $\tau_{R\phi} > 0$, angular momentum is transported outward, and we assume the value of $\nu_\text{disk}$ is determined by accretion disk physics. We treat the value of $\nu_\text{BL}$ as a free parameter, which is a fraction of $\nu_\text{disk}$.

The viscous runs span the full range of azimuthal angle, $0 < \phi < 2 \pi$, and have radial extent $.7 < R/R_* < 12$ with logarithmic scaling in the radial direction. The grid dimensions of the viscous runs are $N_R \times N_\phi = 1024 \times 1024$. As in our MHD runs, we use an isothermal equation of state in our 2D viscous runs, but with $c_s = 1/6$. We use the same initial rotation profile and density profiles as in the unstratified MHD simulation (equations (\ref{eq:rot}) and (\ref{eq:rho})). In particular, the surface density in the disk is initially constant, $\Sigma = 1$ for $R/R_*>1$. On top of the background equilibrium state, we seed the simulations with random perturbations to the initial density. These perturbations trigger shear-acoustic instabilities in the BL, similar to the 3D MHD run. 

In all our viscous simulations, we set $\nu_\text{disk} = 5 \times 10^{-4}$. If we take $H=c_s/\Omega$  in equation (\ref{eq:nu_disk}), this corresponds to an $\alpha$-parameter value of $\alpha \sim .02$ in the inner disk. The only parameter we vary in the viscous simulations is the ratio of the viscosity in the BL to the viscosity in the disk $\nu_\text{BL}/\nu_\text{disk}$. In particular, we present the results of two simulation runs: one with $\nu_{BL}/\nu_\text{disk} = 0.001$ and one with $\nu_{BL} = 0$.

The advantage of the viscous runs is that we know the steady state solution in the disk. This solution for the disk structure should approximately apply even if mass piles up in the BL, because in viscous theory the mass accretion rate, 
\begin{align}
\label{Mdisk}
\dot{M}_\text{disk} = -2 \pi R \Sigma v_R, 
\end{align} 
is set at the outer edge of the disk. The sum of the advected and stress angular momentum currents is also constant\footnote{Unlike $\dot{J}_\text{disk}$, $C_L$ contains only the stress component of the angular momentum current, not the advected component.}:
\begin{align}
\dot{J}_\text{disk} &\equiv -2 \pi R^3 \Omega \Sigma v_R + 2 \pi R^3 \nu \Sigma \frac{d \Omega}{d R} \\
\label{Jdisk}
&= \dot{M}_\text{disk} R^2 \Omega + 2 \pi R^3 \nu_\text{disk} \Sigma \frac{d \Omega}{d R}.
\end{align}
The value of the constant $\dot{J}_\text{disk}$ is set at the inner edge of the disk where the rotation profile turns over ($d\Omega/dR = 0$) and the viscous stress vanishes. For a radially thin BL, we have
\begin{align}
\dot{J}_\text{disk} \approx \dot{M}_\text{disk} R_*^2 \Omega_K(R_*).
\label{Jdiskapprox}
\end{align}
Note that a positive value of $\dot{J}_\text{disk}$ means that the star is gaining angular momentum from the disk.

We can solve for the steady state value of $v_R$ in terms of $\nu_\text{disk}$ by rearranging equation (\ref{Jdisk}):
\begin{align}
v_R = - \frac{\dot{J}_\text{disk}}{2 \pi R^3 \Omega \Sigma} + \frac{\nu}{R} \frac{d \ln \Omega}{d \ln R}.
\label{vrJeq}
\end{align}
For $R \gg R_*$ the first term vanishes, and for a Keplerian rotation profile the mass accretion rate is
\begin{align}
\dot{M}_\text{disk} &\approx 3 \pi \Sigma \nu_\text{disk}, \ \ \ R \gg R_*.
\label{Mdiskapprox}
\end{align}
This motivates us to initialize the velocity profile as
\begin{align}
v_R(R) = \begin{cases} 
      0 & R < 1 \\
      -3\nu_\text{disk}/2R & R\geq 1. 
   \end{cases},
\label{eq:vel_init}
\end{align}

\subsection{Verification}

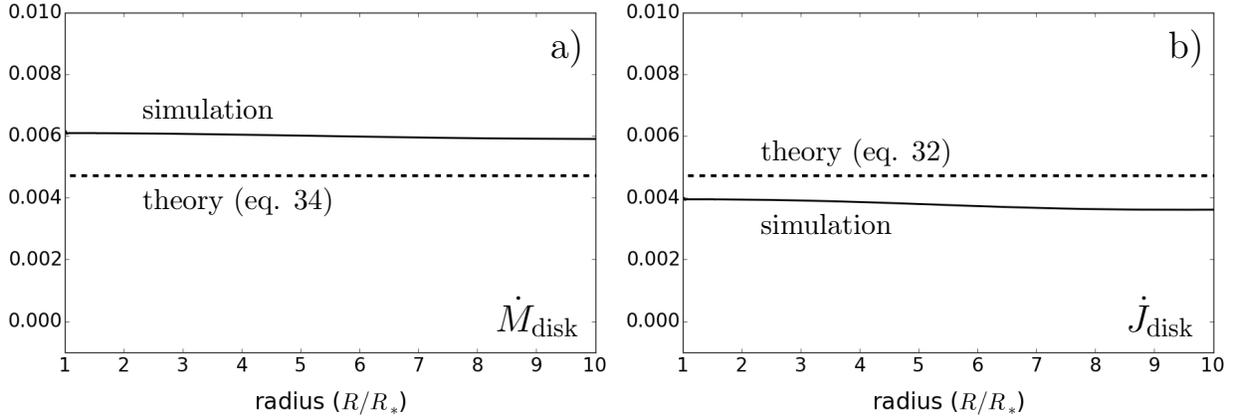
\begin{figure}[!t]
\centering
\subfigure{\begin{overpic}
	[width=0.49\textwidth]{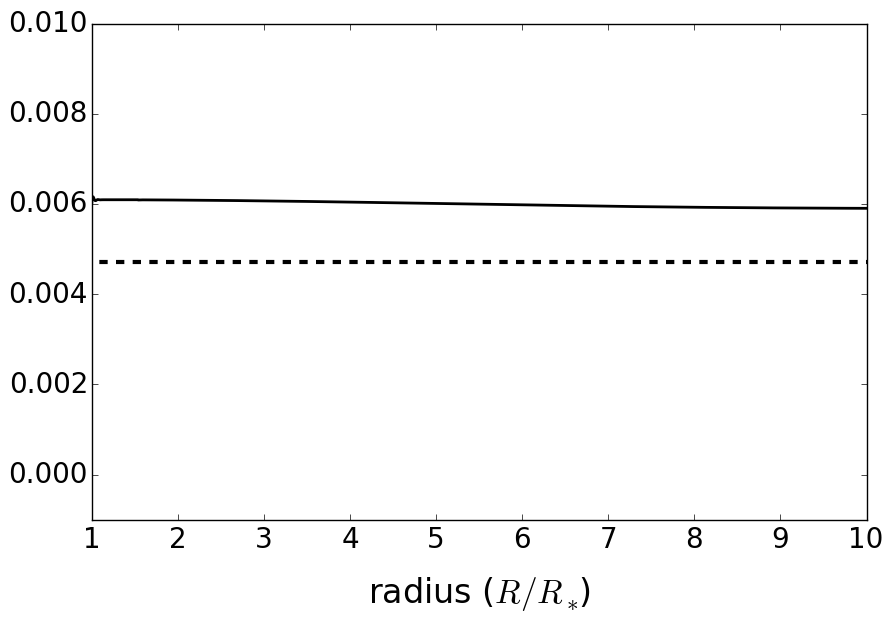}
	\put(81,15){\Large $\dot{M}_\text{disk}$}
	\put(90,60){\Large a)}
	\put(23,35){theory (eq. \ref{Mdiskapprox})}
	\put(23,50){simulation}
\end{overpic}}
\subfigure{\begin{overpic}
	[width=0.49\textwidth]{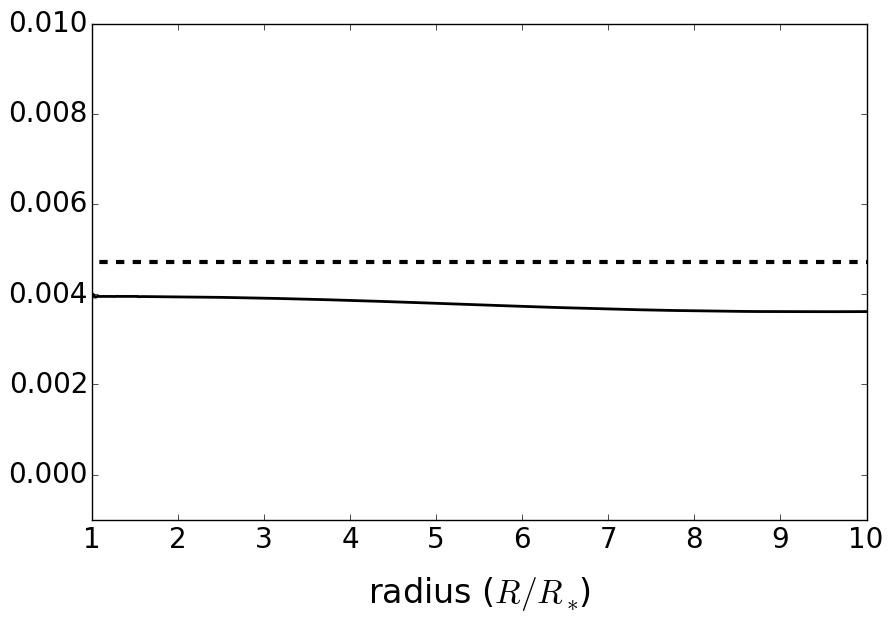}
	\put(83,15){\Large $\dot{J}_\text{disk}$}
	\put(90,60){\Large b)}
	\put(23,43){theory (eq. \ref{Jdiskapprox})}
	\put(23,31){simulation}
\end{overpic}}
\caption{2D viscous hydro simulation with $\nu_\text{BL}$ = 0. Panel a: Mass accretion rate as a function of radius in the disk. Solid curve is the mass accretion rate in the simulation at $t=3000 P_{K,*}$, and dashed curve is the theoretical value using equation (\ref{Mdiskapprox}). The discrepancy between the solid and dashed curves is due to the breakdown of the approximation $R \gg R_*$ in equation (\ref{Mdiskapprox}). A better approximation is to use equation (\ref{vrJeq}) for the radial velocity and set the radius to the outer radius of the simulation domain: $R=R_\text{max}=12R_*$. This gives a theoretical estimate of $\dot{M}_\text{disk} = .00607$, in good agreement with the simulation. Panel b: Solid curve is the sum of the advective and viscous angular momentum currents in the simulation according to equation (\ref{Jdisk}) in the disk at $t=3000 P_{K,*}$. In steady state, this sum is a constant which determines the value of $\dot{J}_\text{disk}$. The dotted line shows the theoretical estimate for $\dot{J}_\text{disk}$ according to equation (\ref{Jdiskapprox}). The solid curve lies below the dashed line, because the angular velocity is sub-Keplerian in the inner part of the disk. }
\label{steadytestfig}
\end{figure}

The velocity profile in equation (\ref{eq:vel_init}) gives the correct steady state mass accretion rate in the outer disk ($R \gg R_*$) for a constant surface density profile. It would give the exact steady state solution at all radii in the disk if $\dot{J}_\text{disk} = 0$. However the value of $\dot{J}_\text{disk}$ is set at the inner edge of the accretion disk and is given by equation (\ref{Jdiskapprox}) for a slowly-rotating star. Thus, the disk density, radial velocity, and (to a lesser extent) angular velocity will readjust until the correct steady state value of $\dot{J}_\text{disk}$ is established throughout the disk. This readjustment happens on a viscous timescale, starting from the BL and proceeding outwards through the disk. 

The steady state values of $\dot{M}_\text{disk}$ and $\dot{J}_\text{disk}$ provide a check of our viscous simulations at late times. The solid black curves in Figs.\ \ref{steadytestfig}a,b show $\dot{M}_\text{disk}$ and $\dot{J}_\text{disk}$, respectively, as a function of radius in the 2D viscous hydro simulation with $\nu_\text{BL}$ = 0. The curves are plotted at the time $t=3000 P_{K,*}$ when the disk in the simulation is close to steady state. The dashed lines show the theoretically-predicted values for $\dot{M}_\text{disk}$ and $\dot{J}_\text{disk}$, respectively.

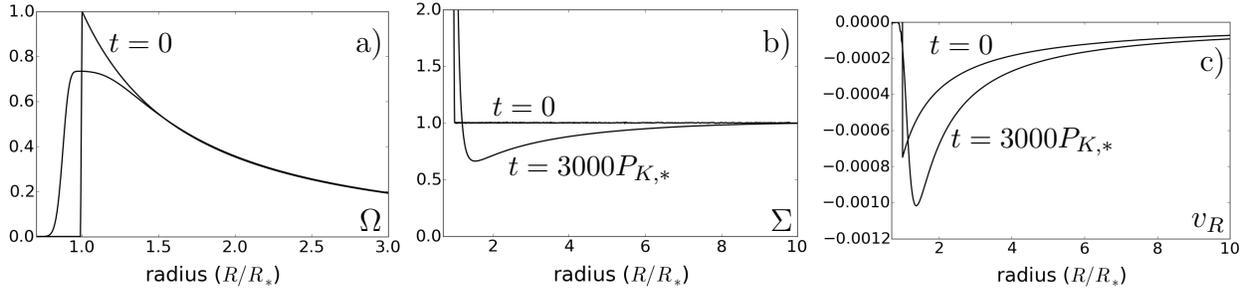
\begin{figure}[!t]
\centering
\subfigure{\begin{overpic}
	[width=0.32\textwidth]{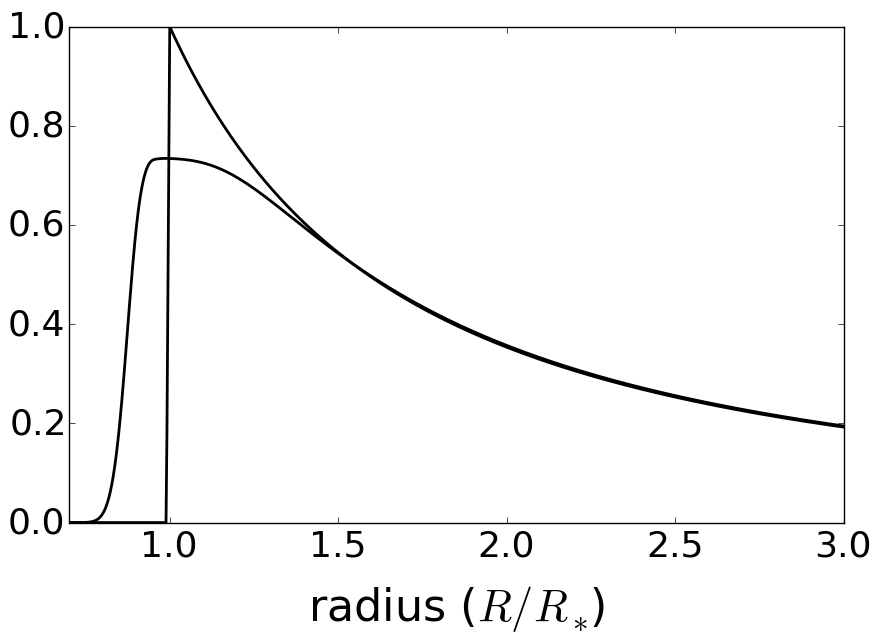}
	\put(89,15){\large $\Omega$}
	\put(88,60){\large a)}
	\put(26,60){$t=0$}
\end{overpic}}
\subfigure{\begin{overpic}
	[width=0.32\textwidth]{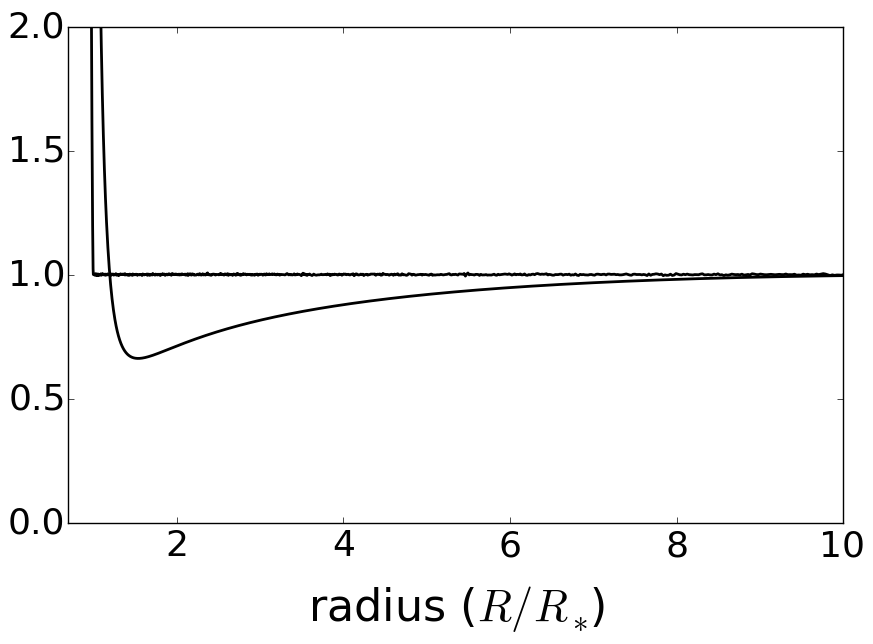}
	\put(90,15){\large $\Sigma$}
	\put(88,60){\large b)}
	\put(20,44){$t=0$}
	\put(24,29){$t=3000 P_{K,*}$}
\end{overpic}}
\subfigure{\begin{overpic}
	[width=0.34\textwidth]{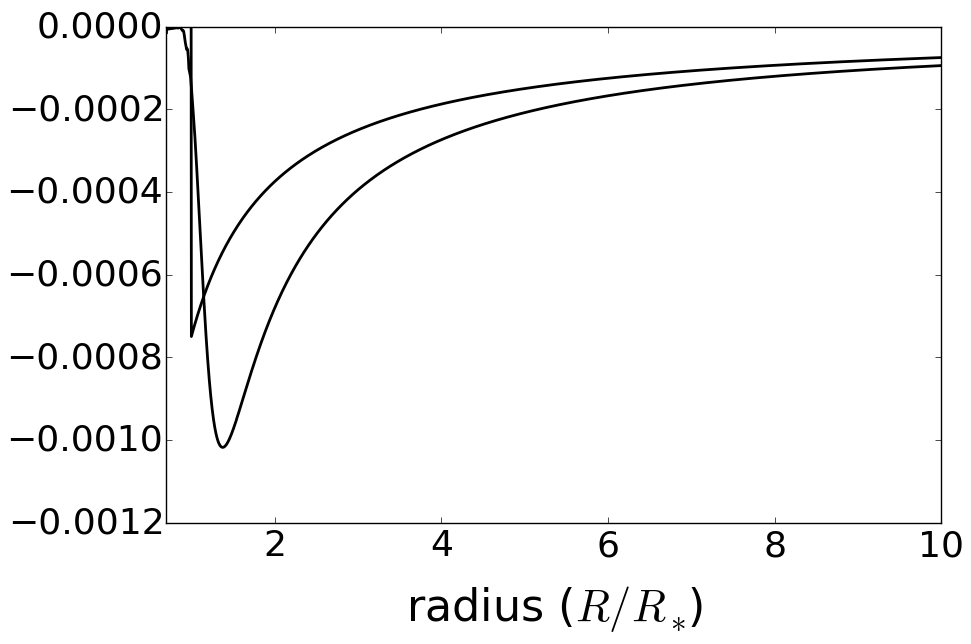}
	\put(88,15){\large $v_R$}
	\put(90,52){\large c)}
	\put(26,56){$t=0$}
	\put(31,34){$t=3000P_{K,*}$}
\end{overpic}}
\caption{2D viscous hydro simulation with $\nu_\text{BL}$ = 0.  Panel a: Angular velocity profile in the simulation at $t=0$ (labeled curve) and at $t=3000 P_{K,*}$ (unlabeled curve). Panel b: Surface density profile at $t=0$ and at $t=3000 P_{K,*}$. Panel c: Radial velocity at $t=0$ and at $t=3000 P_{K,*}$.}
\label{vistestfig}
\end{figure}

Panels a, b and c of Fig.\ \ref{vistestfig} show the angular velocity, surface density, and radial velocity profiles, respectively, at $t=0$ and at $t=3000P_{K,*}$ for the 2D viscous hydro simulation with $\nu_\text{BL}$ = 0. In panel c, the radial velocity is more negative in the inner part of the disk at $t=3000P_{K,*}$ compared to $t = 0$. This should be the case according to equation (\ref{vrJeq}), because $\dot{J}_\text{disk} = 0$ initially, but then readjusts to its positive steady state value at late times. The drop in surface density in the inner part of the disk is explained by the somewhat larger (in magnitude) velocity in that region together with the requirement that the mass accretion rate through the disk should be constant in steady state.

\subsection{Angular Momentum Belt}

In \S \ref{sec:belt}, we saw that in the 3D MHD simulation acoustic waves excited in the BL were not enough to transport angular momentum advected into the BL from the disk. We may ask whether this also holds for 2D viscous hydro simulations? We begin by discussing the 2D viscous simulation which has $\nu_\text{BL} = 0$, and thus no viscous transport of angular momentum radially inward of the point where $\tau_{R\phi} = 0$. In the absence of any transport mechanism except viscosity, accreted material would pile up in the BL. Shear-acoustic instabilities are still excited in 2D viscous simulations, but are they enough to stave off accumulation of angular momentum in the BL?

\begin{figure}[!t]
\centering
\subfigure{\begin{overpic}
	[width=0.5\textwidth]{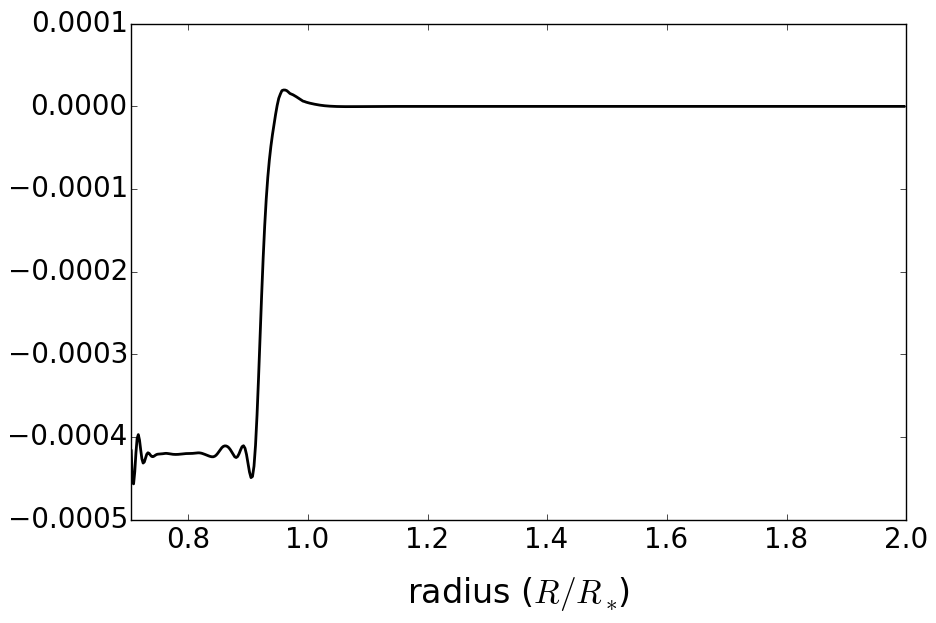}
	\put(82,15){\Large $C_{L,H}$}
\end{overpic}}
\subfigure{\begin{overpic}
	[width=0.49\textwidth]{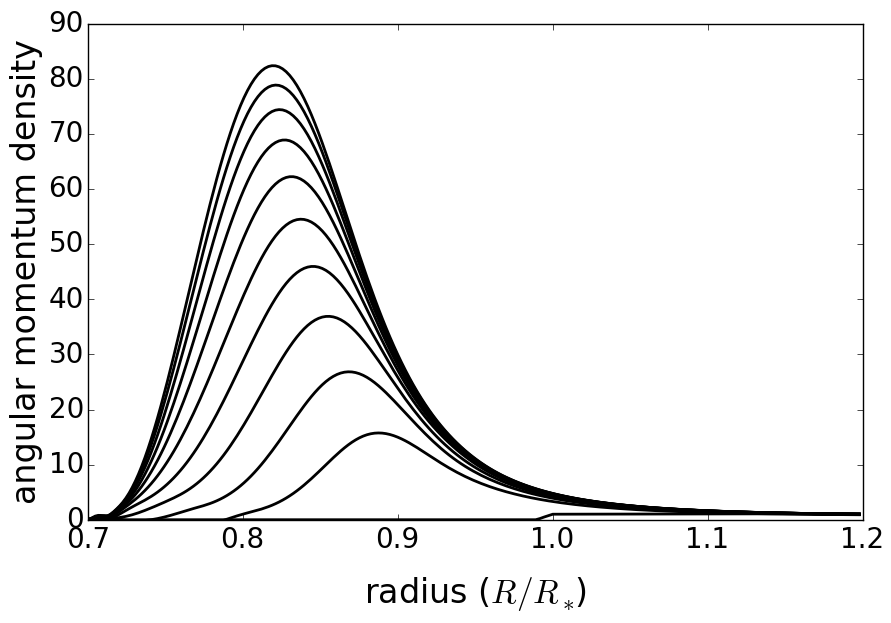}
	\put(34,61){\small $t=3000 P_{K,*}$}
	\put(80,59){\Large $\rho R^2 \Omega$}
\end{overpic}}
\caption{2D viscous hydro simulation with $\nu_\text{BL}=0$. Left panel: The hydrodynamical angular momentum current, $C_{L,H}$, at $t=3000P_{K,*}$. $C_{L,H}$ is negative in the star due to sound waves excited in the BL that transport angular momentum into the star. Right panel: angular momentum density ($\rho R^2 \Omega$) in intervals of $\Delta t = 300 P_{K,*}$ from $t=0$ to $t=3000 P_{K,*}$. A belt of angular momentum forms in the BL, the amplitude of which grows monotonically. This is very similar to what we found in the 3D MHD simulation (Fig. \ref{fig:rot}). }
\label{visstressfig}
\end{figure}

The left panel of Fig.\ \ref{visstressfig} shows the hydrodynamical stress angular momentum current, $C_{L,H}$, in the star and the BL for the 2D viscous simulation with $\nu_{BL} = 0$ at $t=3000P_{K,*}$. $C_{L,H}$ inside the star is relatively constant and negative, meaning waves do transport some of the accreted angular momentum radially inward. However, comparing the left panel of Fig.\ \ref{visstressfig} with Fig.\ \ref{steadytestfig}b, $-C_{L,H}/\dot{J}_\text{disk} \sim 0.1$. Therefore, waves in the star transport angular momentum away from the BL at a rate that is only about 10\% of the rate it is transported into the BL from the disk. This value of 10\% is also consistent with our estimate for the 3D MHD simulation. 

One may wonder whether advection can carry the angular momentum inside the star instead of waves? However, examining equation (\ref{Jdiskapprox}), this is impossible in steady state ($\dot{M}_\text{disk} = \dot{M}_\text{star}$) for a slowly rotating star ($\Omega_* \ll \Omega_K(R_*)$).

If neither acoustic waves excited in the BL nor advection can effectively transport angular momentum in the star, angular momentum will pile up in the BL. The right panel of Fig.\ \ref{visstressfig} shows a plot of the angular momentum in the simulation with $\nu_\text{BL} = 0$ at different times. As expected angular momentum accumulates in the BL, forming a rapidly rotating belt. Moreover, this pile up continues for the duration of the simulation ($\approx 3000 P_{K,*}$) with no sign of stopping. The angular momentum evolution in the 2D hydro simulation in Fig.\ \ref{visstressfig} (right panel) is strikingly similar to that in the 3D MHD simulation in Fig.\ \ref{fig:rot} (right panel).

\begin{figure}[!t]
\centering
\subfigure{\begin{overpic}
	[width=0.5\textwidth]{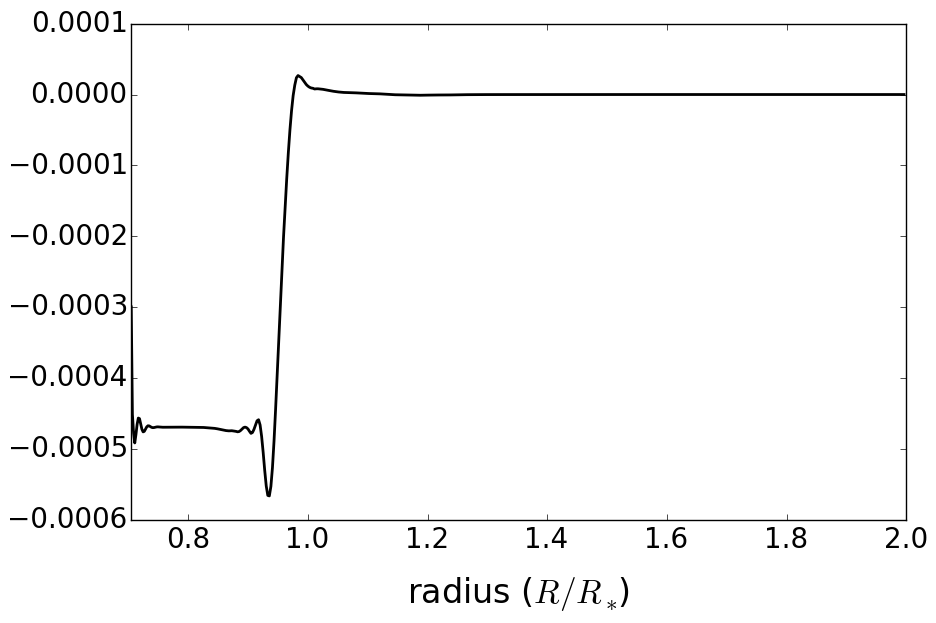}
	\put(82,15){\Large $C_{L,H}$}
\end{overpic}}
\subfigure{\begin{overpic}
	[width=0.49\textwidth]{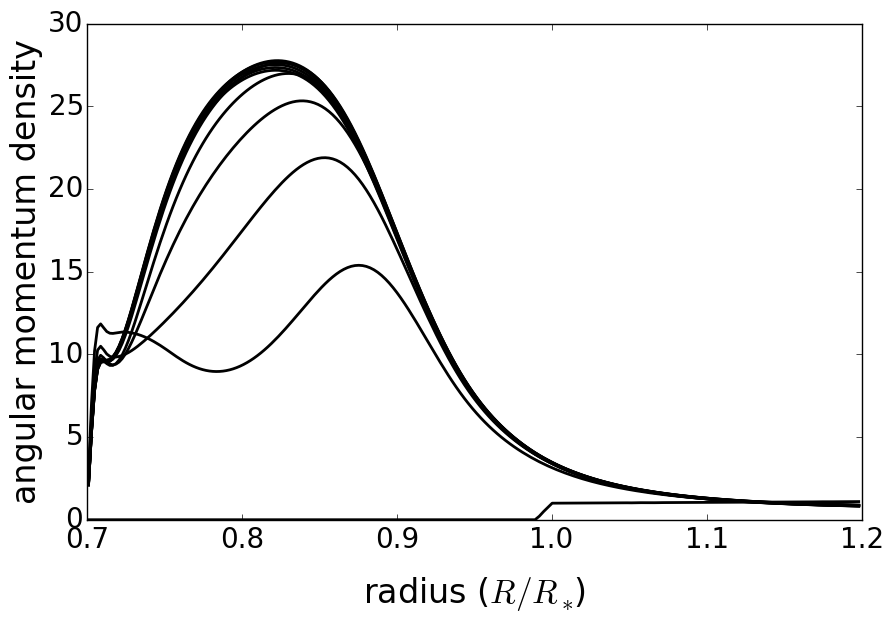}
	\put(37,61){\small steady state}
	\put(50,12){\small $t=0$}
	\put(18,24){\small $t=300P_{K,*}$}
	\put(80,59){\Large $\rho R^2 \Omega$}
\end{overpic}}
\caption{2D viscous hydro simulation with $\nu_\text{BL}/\nu_\text{disk}=0.001$. Left panel: The hydrodynamical angular momentum current, $C_{L,H}$ at $t=3000P_{K,*}$. $C_{L,H}$ is negative in the star and of similar amplitude as the $\nu_\text{BL} = 0$ case (left panel of Fig.\ \ref{visstressfig}). Right panel: angular momentum density ($\rho R^2 \Omega$) in intervals of $\Delta t = 300 P_{K,*}$ from $t=0$ to $t=3000 P_{K,*}$. A belt of angular momentum does begin to form in the BL. However, its amplitude saturates, as the simulation approaches steady state after $t \approx 1500 P_{K,*}$. The small but non-zero viscosity in the BL ($\nu_\text{BL}$) dominates over the wave angular momentum transport in the BL.}
\label{visstressfig1}
\end{figure}
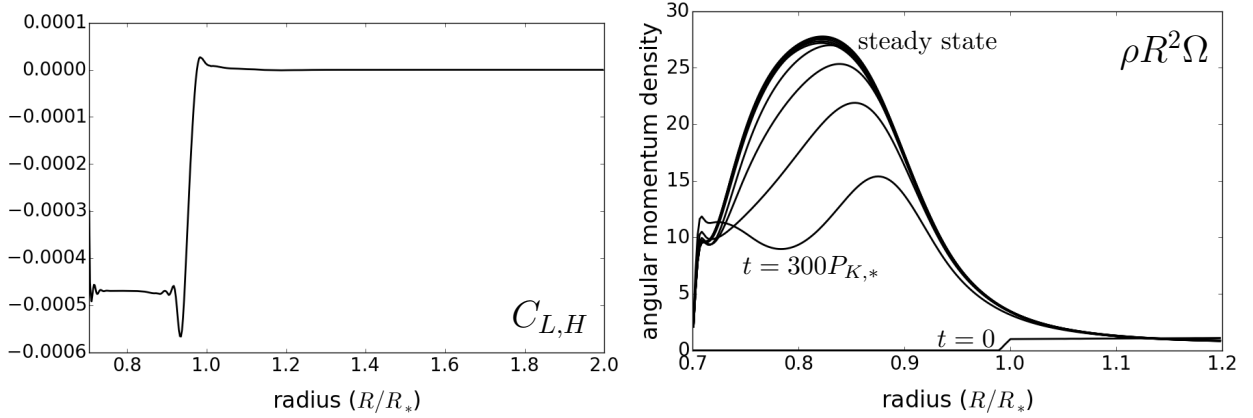

Next, we show how the presence of some viscosity in the BL affects the accumulation of angular momentum there. Fig.\ \ref{visstressfig1} is the same as Fig.\ \ref{visstressfig}, but for $\nu_\text{BL}/\nu_\text{disk}=.001$. The left panel of Fig.\ \ref{visstressfig1} shows the hydrodynamical stress, $C_{L,H}$, at $t=3000P_{K,*}$. The angular momentum current in the star is similar to the case of $\nu_\text{BL}=0$, suggesting that angular momentum transport due to waves is at a similar level in both simulations. However, unlike the case of $\nu_\text{BL} = 0$, the simulation with $\nu_\text{BL}/\nu_\text{disk}=.001$ does reach a steady state. The right panel of Fig.\ \ref{visstressfig1} shows the angular momentum density. A belt still forms in the BL, as before, but the amplitude of the belt saturates around $t\approx1500P_{K,*}$. Note that most of the angular momentum in steady state is carried into the star and the BL by the small explicit viscosity, not by acoustic waves. The waves again carry only about $\sim 10\%$ of the total angular momentum required for steady state accretion.

\subsection{Time to Reach Steady State}
\label{sec:visc_steady}
We can understand many of the features of the viscous simulations using dimensional analysis. The time for the BL to reach steady state is of order the viscous time in the BL:
\begin{align}
\tau_\text{BL} \sim \frac{\delta_\text{BL}^2}{\nu_\text{BL}},
\label{eq:tauBL}
\end{align}
where $\delta_\text{BL}$ is the steady state dynamical width of the BL. This is the radial extent over which the angular velocity adjusts from its Keplerian value in the disk, $\Omega_K(R_*)$, to its stellar value, $\Omega_*$. 

To estimate the value of $\delta_\text{BL}$, we can employ an argument first used by \cite{PringleBL}. In steady state, the radial momentum equation can be written as
\begin{align} 
\rho v_R \frac{d v_R}{d R} = - \frac{dP}{dR} - \rho (\Omega_K^2-\Omega^2)R,
\label{rad_mom}
\end{align}
where $\Omega(R)$ is the 1D angular velocity profile. Setting $P \sim \rho c_s^2$, where $c_s$ is a characteristic sound speed in the BL, we see that the ratio of the term on the left and the first term on the right of equation (\ref{rad_mom}) is
\begin{align}
\left|\rho v_R \frac{d v_R}{d R} \right| \times \left|\frac{dP}{dR}\right|^{-1} \sim \left(\frac{v_R}{c_s}\right)^2. 
\end{align}
As we have already remarked, the value of $\dot{J}_\text{disk}$ is set at the inner boundary of the disk in steady state. Therefore, in order for steady state disk theory to apply, the inflow velocity to the BL must be subsonic. As a result, $v_R/c_s < 1$, and we can equate the two terms on the right hand side of equation (\ref{rad_mom}) to estimate the width of the BL:
\begin{align}
\delta_\text{BL} = f_\text{BL} R_* \left(\frac{c_s}{V_K(R_*)}\right)^2.
\label{eq:deltaBL}
\end{align}
Here $f_\text{BL} \gtrsim 1$ is a dimensionless constant, and the width of the BL scales with the radial pressure scale height in the star (equation (\ref{eq:scale})).

Defining the Mach number in the BL as 
\ba
\mathcal{M}_\text{BL} \equiv \frac{V_K(R_*)}{c_s},
\ea
and plugging the BL width from equation (\ref{eq:deltaBL}) into equation (\ref{eq:tauBL}), the viscous time in the BL is
\begin{align}
\tau_\text{BL} \sim \frac{f_\text{BL}^2 R_*^2}{\nu_\text{BL} \mathcal{M}_\text{BL}^4}.
\label{tau_visc_BL}
\end{align}
We can check this formula against the 2D viscous simulation with $\nu_\text{BL}/\nu_\text{disk}=.001$. From the right panel of Fig.\ \ref{visstressfig1}b, the time for the simulation to reach steady state is $t \sim 1500 P_{K,*}$. Setting $\mathcal{M}=6$, $R_* = 1$, and $\nu_\text{BL} = 5 \times 10^{-7}$, as appropriate for that simulation, we have that the viscous time in the BL is $\tau_\text{BL} \sim 300 f_\text{BL}^2 P_{K,*}$ from equation (\ref{tau_visc_BL}).  Setting the viscous time equal to the time required to reach steady state implies that $f_\text{BL} \sim 2-3$. According to equation (\ref{eq:deltaBL}), this value of $f_\text{BL}$ implies $\delta_\text{BL} \sim .07-.1$, which is consistent with the BL width in the simulation.  

\subsection{Condition for Belt Formation}

Next, we derive a condition for the formation of an angular momentum belt in the BL and estimate its amplitude. If we assume the angular momentum current in the BL is predominantly carried by viscous stresses rather than waves, we can write
\begin{align}
2\pi \nu_\text{BL} \Sigma_\text{BL} R^3 \frac{d \Omega}{dR} \sim \dot{J}_\text{disk},
\label{eq:JdotBL}
\end{align}
where $\Sigma_\text{BL}$ is a characteristic density that we intend to solve for. In equating the left and right sides of equation (\ref{eq:JdotBL}), we have used the fact that $\dot{J}$ must take the same constant value everywhere (i.e.\ in the star, the disk, and the BL) in steady state. Substituting the value of $\dot{J}_\text{disk}$ from equation (\ref{Jdiskapprox}) into equation (\ref{eq:JdotBL}) and dropping constants of order unity we can estimate
\begin{align}
\Sigma_\text{BL} &\sim \Sigma_\text{disk} \left(\frac{\delta_\text{BL}}{R_*}\right) \left(\frac{\nu_\text{disk}}{\nu_\text{BL}}\right) \\
& \sim \Sigma_\text{disk} f_\text{BL} \mathcal{M}_\text{BL}^{-2} \left(\frac{\nu_\text{disk}}{\nu_\text{BL}}\right).
\label{eq:sigmaBL}
\end{align}

We parametrize the characteristic value of the angular momentum density in the BL as
\begin{align}
L_\text{BL} = l_\text{BL} \Sigma_\text{BL} \Omega_K(R_*) R_*^2,
\label{eq:lmax}
\end{align}
where $l_\text{BL} \lesssim 1$ is a dimensionless constant. A belt of angular momentum will exist in the BL if $ \Sigma_\text{BL} l_\text{BL} \Omega_K(R_*) R_*^2 \gg \Sigma_\text{disk} \Omega_K(R_*) R_*^2$. Using equation (\ref{eq:sigmaBL}), we can write the condition for belt formation as  
\begin{align}
f_\text{BL} l_\text{BL} \mathcal{M}_\text{BL}^{-2} \left(\frac{\nu_\text{disk}}{\nu_\text{BL}}\right) \gg 1.
\label{eq:belt_cond}
\end{align}

In the viscous simulation with $\nu_\text{BL}/\nu_\text{disk}=.001$, the condition expressed in equation (\ref{eq:belt_cond}) is satisfied. Using equation (\ref{eq:sigmaBL}) for $\Sigma_\text{BL}$, equation (\ref{eq:lmax}) predicts $L_\text{BL} \sim 30 l_\text{BL} f_\text{BL} \Sigma_\text{disk} \Omega_K(R_*) R_*^2$. Since $\Sigma_\text{disk} \Omega_K(R_*) R_*^2 \sim 1$ in our units, this estimate is a good match to the maximum steady state value of the angular momentum in the belt within the simulation (right panel of Fig.\ \ref{visstressfig1}) if $l_\text{BL}f_\text{BL} \sim 1$.   

\subsection{Implication for Viscous Models of the BL}

The condition in equation (\ref{eq:belt_cond}) is not trivial to satisfy, since $\mathcal{M}_\text{BL} \ll 1$. Therefore, we may ask whether we expect an angular momentum belt to form in published viscous models of the BL \citep{PophamNarayan,Kley,HertfelderKley1}?

\cite{PophamNarayan} argued that because the radial pressure scale height in the star is smaller than the vertical scale height in the BL by a factor of $\mathcal{M}_\text{BL}$, the viscosity should be parametrized as
\begin{align}
\nu = \alpha c_s \text{min}\left(h_{R,*},H\right).
\label{eq:nuall}
\end{align} 
Here $h_{R,*}$ (equation (\ref{eq:scale})) and $H$ are the radial pressure scale height in the star and the vertical scale height in the disk, respectively. Equation (\ref{eq:nuall}) is a reasonable physical ansatz that can be used in the star, the BL, and the disk. Moreover, one can show that it leads to subsonic radial inflow through the BL for $\alpha \ll 1$. However, a criticism of the ansatz is that it assumes $\alpha_\text{BL}$ = $\alpha_\text{disk}$ which is not necessarily true given that the physical mechanisms leading to angular momentum transport in the BL and the disk are different.

Nevertheless, we may ask whether a BL solution employing the ansatz in equation (\ref{eq:nuall}) forms a belt of angular momentum in the BL?
Taking 
\begin{align}
\nu_\text{BL} = \alpha c_s h_{R,*},
\label{eq:nuBL}
\end{align} 
and substituting equation (\ref{eq:nuBL}) into equation (\ref{eq:belt_cond}) the condition for angular momentum belt formation within this viscosity model is
\begin{align}
f_\text{BL} l_\text{BL} \mathcal{M}_\text{BL}^{-1} \gg 1.
\end{align}
Since $\mathcal{M}_\text{BL} \gg 1$ and $f_\text{BL} l_\text{BL} \sim 1$, the condition is not met and a belt of angular momentum does not form. The fact that an angular momentum belt {\it does} form in our 3D MHD simulations means that the viscosity in the BL in these simulations is much smaller than what is predicted by the ansatz in equation (\ref{eq:nuall}). 

Because the amplitude of the angular momentum belt in the BL grows without bound in the 3D MHD simulation, we cannot explicitly compute the effective turbulent viscosity in the BL in the simulation. However, we can place an upper bound on it. Using equation (\ref{tau_visc_BL}), we can solve for the viscosity in the BL in terms of the viscous time. Taking the viscous time in the BL equal to the duration of our 3D MHD simulation ($\tau_\text{BL} = 936 P_{K,*}$), we can set a lower bound to the effective value of the turbulent viscosity in the BL in the simulation: 
\begin{align}
\nu_\text{BL} &= \frac{f_\text{BL}^2 R_*^2}{\tau_\text{BL} \mathcal{M}_\text{BL}^4} \\
&< 10^{-7} \left(\frac{f_\text{BL}}{2.5}\right)^2.
\label{nuBLmhd}
\end{align}
Here the only uncertainty is in the dimensionless parameter $f_\text{BL}$ which parametrizes the width of the boundary layer in terms of the number of scale heights. We have used a value of $f_\text{BL} = 2.5$ based on the results of viscous simulations (\S \ref{sec:visc_steady}).

Given the estimate of $\nu_\text{BL}$ in equation (\ref{nuBLmhd}), we can derive an upper bound to the value of $\alpha$ in the BL. Starting from equation (\ref{eq:nuBL}), we can write
\begin{align}
\alpha_\text{BL} &= \frac{\nu_\text{BL}}{c_s h_{R,*}} < 10^{-4}.
\label{alphabound}
\end{align}

The value of $\alpha_\text{disk}$ in the 3D MHD simulation due to MRI turbulence in the disk was found to be $\alpha_\text{disk} \sim .01-.05$ (\S \ref{sec:verify}). Thus, $\alpha_\text{BL} \ll \alpha_\text{disk}$, {\it even though} equation (\ref{alphabound}) only gives an upper bound on the value of $\alpha_\text{BL}$.

\section{Discussion}
\label{sec:discuss}

We have shown using 3D MHD simulations that a belt of angular momentum forms in the boundary layer as a result of accretion driven by MRI in the disk onto the surface of a star. The belt of angular momentum grows in amplitude without bound over the course of $\sim$1000 Keplerian orbital periods at the inner edge of the disk. This implies that there is not enough angular momentum transport in the BL within our simulations to carry all of the angular momentum of the accreted material into the star. 

This is in spite of the fact that accretion advects magnetic field generated by MRI turbulence in the disk into the BL. In particular, we do not see significant amplification of magnetic field in the BL, which contradicts \cite{Armitage} who claimed magnetic activity in the BL. However, because he initialized the disk with a net vertical flux, the accumulation of magnetic field he observed in the BL may be due to flux dragging and the frozen-in-law, just as in our MHD simulation (see Figs.\ \ref{fig:flux} \& \ref{fig:flux2}). \cite{Armitage} also did not provide plots of $B_R$ or $B_\phi$ which would have supported the claim of magnetic field amplification. Our results are in line, though, with \cite{PessahChan} who showed that although the energy density of sheared magnetic waves can be amplified by an order of magnitude in the BL, the stresses due to these waves oscillate around zero. In the future, it would be interesting to investigate whether the transient spikes observed in the $\phi$-component of the magnetic energy density in our MHD simulation (panel d of Fig.\ \ref{fig:flux2}) are related to the swing amplification mechanism studied by \cite{PessahChan}.
 
Inefficient angular momentum transport in the BL within our simulations is particularly puzzling given that shear-acoustic instabilities are excited in the BL and persist for the duration of each simulation. As a result, the hypothesis of \cite{BRS1,BRS2} that waves efficiently transport angular momentum in the BL appears to be invalid. In particular, \cite{BRS2} envisaged that the outbursts of shear-acoustic instability in the BL result in a limit cycle behavior that regulates the flow of material through the BL. However, the simulations of \cite{BRS2} were run for only $\sim 100-200 P_{K,*}$, whereas our 3D MHD simulation is run for almost $1000 P_{K,*}$. On these longer timescales, we do not see the limit cycle behavior continuing after the two large outbursts of shear-acoustic instability around $t=70P_{K,*}$ and $t=220P_{K,*}$, as seen in the bottom panel of Fig.\ \ref{fig:stress} and in Fig.\ \ref{dens_st_fig}. Moreover, we find that waves only carry a fraction of the angular momentum required to achieve steady state ($\sim 10\%$) within the star and the BL at late times in our simulations.

We also ran 2D viscous hydro simulations for which we could control the steady state mass accretion and angular momentum transport rates through the disk. These simulations confirmed that a rapidly rotating belt of accreted material forms in the BL, because of inefficient transport of angular momentum through the BL and the star. Using dimensional analysis, we were able to show that when the viscosity in the BL falls below a critical value, a belt of angular momentum forms in the BL (equation (\ref{eq:belt_cond})). However, as long as the viscosity in the BL is greater than zero, the amplitude of the angular momentum belt eventually saturates at a value given approximately by equation (\ref{eq:lmax}).

If waves are insufficient to transport the angular momentum in the BL, then it seems we must fall back on viscosity to do the job. However,
our 3D MHD results combined with our 2D viscous hydro results suggest that the viscous coupling between the star and the accretion disk via the BL is much weaker than is typically assumed in viscous models of the BL that use an ansatz like the one in equation (\ref{eq:nuall}). This is important, because viscous models are still the standard way of connecting BL theory with observations.

 Consequently, we believe it is interesting and physically well-motivated to use different values of $\alpha$ when $\tau_{R,\phi} > 0$ and when $\tau_{R,\phi} < 0$ (i.e.\ in the BL and in the disk). Even though this does not fit neatly into the ansatz of equation (\ref{eq:nuall}), it is supported by the 3D MHD simulations and could lead to more physical models of the BL (e.g.\ ones that contain an angular momentum belt). We point out that there are indications of an angular momentum belt forming in Fig. 8 of \cite{HertfelderKley} due to the flattening in time of the azimuthal velocity profile around the stellar surface (compare with the left panel of our Fig. \ref{fig:rot}).

If global shear instabilities are ineffective at transporting angular momentum in the supersonic regime, then a different mechanism or instability must be responsible. For instance, the Tayler-Spruit dynamo \citep{TSdynamo} is a physical pathway leading to turbulence that could provide an effective viscosity in the BL where $d \Omega/dr > 0$. However, the efficiency of this transport process is still not well understood theoretically, nor are the resolution requirements for capturing it numerically \citep{IB15}. Another possible instability that could drive angular momentum transport is baroclinic instability. However, determining if baroclinic instability is important for angular momentum transport in the BL would require stratification in the $z$-direction and an accurate model of BL thermodynamics. 

We can make some general statements regarding how the effective temperature of the BL would change in the presence of inefficient angular momentum transport. The effective blackbody temperature of an optically thick BL in steady state can be estimated by considering the luminosity of the BL and the radiating area:
\begin{align}
T_\text{BL} \equiv \left(\frac{\mathcal{L}_\text{BL}}{\sigma A_\text{BL}}\right)^{1/4}.
\end{align}

For a Keplerian disk around a slowly-rotating star, as much kinetic energy remains to be dissipated at the surface of the star as in coming from the outer part of the disk ($R \gg R_*$) to the surface. Therefore, the luminosity of the BL equals the luminosity of the accretion disk $\mathcal{L}_\text{BL} \approx \mathcal{L}_\text{disk}$. In classical BL theory, the radiating area of the BL, on the the hand, is set by the vertical extent of the BL \citep{PringleBL,PophamNarayan}. Up to constant factors of order unity, this equals the disk scale height $H \sim R_* c_s/V_K(R_*) = R/\mathcal{M}_\text{BL}$, where $V_K(R_*)$ is the Keplerian velocity at the surface of the star, $c_s$ is the effective sound speed in the BL, and $\mathcal{M}_\text{BL} \equiv V_K(R_*)/c_s$ is the Mach number in the BL.

However, if angular momentum transport is inefficient and a belt of angular momentum forms, it is possible that this belt will spread latitudinally across the surface of the star, resulting in a spreading layer \citep{InogamovSunyaev,PiroBildsten}. In this case, the radiating area of the BL will increase and the temperature of the BL will decrease. In the extreme case when the belt spreads all the way to the poles, the radiating area will increase by a factor of order $R/H \sim \mathcal{M}_\text{BL}$, and the effective temperature will drop by a factor of $\mathcal{M}_\text{BL}^{1/4}$.

Latitudinal spreading of the belt and the consequent drop in temperature could help to explain white dwarf observations, where the BL appears ``missing" \citep{missingBL,Mukai}. In particular, if there is substantial spreading of the belt across the surface of the star then the temperature of the BL (which is really a spreading layer in this case) will approach the temperature of the inner part of the accretion disk, because the two will emit a comparable amount of power over a comparable amount of surface area. In this scenario, the spectrum of the BL and the inner part of the disk blend together, providing an explanation for the missing BL phenomenon in weakly-magnetized accreting white dwarfs. As a result, the spreading of the angular momentum belt in latitude is an interesting possibility to explore in future work and would allow for a closer connection between dynamical BL theory and observations.

Finally, it is important to understand why shear-acoustic instabilities in the BL are inefficient at transporting angular momentum. One intriguing possibility involves the different physical mechanisms by which shear instabilities operate in the subsonic and supersonic regimes. For instance, consider the Kelvin Helmholtz instability (KHI), which applies to a subsonic jump across a shear layer. KHI can be viewed as a destabilizing interaction between Rossby edge waves on the upper and lower edges of the shear layer \citep{Bretherton,Heifetz}. Because it couples the two edges of the shear layer, it seems reasonable that the nonlinear evolution of the system results in angular momentum transport across the shear layer. 

On the other hand, shear-acoustic instabilities can be viewed as the destabilization of an incompressible mode by direct emission of acoustic radiation \citep{acousticCFS}. This does not involve coupling between the two edges of the shear layer, and the excitation region of an unstable shear-acoustic mode exists over only a small radial extent, near the corotation radius in the BL. Perhaps this radial confinement of the excitation region can help explain our result that waves transport only a small fraction ($\sim 10$\%) of the angular momentum current required to achieve steady state in our simulations. In the future, it would be interesting to study the interaction between inertia-gravity waves and Rossby waves in the BL. For example, excitation of Rossby waves by radiation of inertia-gravity waves (as opposed to acoustic waves) in a vortical shear flow is an important process in meteorology \citep{IGWemit}.

\section*{Acknowledgements}
The authors would like to thank Lars Bildsten, Roman Rafikov, Sasha Philippov, and Bill Wolf for important discussions. This research is funded in part by the Gordon and Betty Moore Foundation through Grant GBMF5076 and by the Simons Foundation through a Simons Investigator Award to EQ. MB was supported by NASA Astrophysics Theory grant NNX14AH49G to the University of California, Berkeley and the Theoretical Astrophysics Center at UC Berkeley. Resources supporting this work were provided by the NASA High-End Computing (HEC) Program through the NASA Advanced Supercomputing (NAS) Division at Ames Research Center.

\end{document}